\documentclass[twocolumn,showpacs,preprintnumbers,amsmath,amssymb,nofootinbib]{revtex4}
\usepackage{graphicx}% Include figure files
\usepackage{dcolumn}% Align table columns on decimal point
\usepackage{bm}% bold math
\usepackage{amssymb}
\usepackage{amsmath}
\usepackage{amsthm}
\begin{document}

%\preprint{APS/123-QED}

\title{Correlation between multiple scattering angle and  ionization energy loss \\
for fast electrons}

\author{M.V.~Bondarenco}
 \email{bon@kipt.kharkov.ua}
%\email{bon@kipt.kharkov.ua}
%\author{N.F.~Shul'ga}
 \affiliation{%
NSC Kharkov Institute of Physics and Technology, 1 Academic St.,
61108 Kharkov, Ukraine }
% \affiliation{%
%V.N. Karazine Kharkov National University, 4 Svobody Sq., 61077 Kharkov, Ukraine }

\date{\today}% It is always \today, today,
             %  but any date may be explicitly specified

\begin{abstract}

A correlation between the angle of multiple scattering and the ionization energy loss for relativistic electrons in an amorphous medium is computed by solving the combined transport equation. 
The correlation is found to be the most pronounced at deflection angles larger than typical, reflecting the underlying single-scattering kinematical correlation,
but is also sizable at typical deflection angles, 
where the width of the angular distribution increases
%may be expressed in terms of effective extension of the particle path length in the target 
with the increase of the energy loss.
%There also remains a strong residual correlation in the Rutherford asymptotic region, but it is smeared by elastic multiple scattering on atomic nuclei, developing a broad correlation ridge. 
The mean energy loss as a function of the deflection angle is calculated. 
It grows quadratically both at small and at large angles, 
but the proportionality coefficient at large angles is greater than at small ones.

%Dimuon registration efficiency is evaluated in analytic form.

\end{abstract}

%% keywords here, in the form: keyword \sep keyword


%% MSC codes here, in the form: \MSC code \sep code
%% or \MSC[2008] code \sep code (2000 is the default)

%\end{keyword}

\maketitle

\section{Introduction}\label{sec:Intro}

Fast charged particles passing through amorphous matter deflect and lose energy by elastic and inelastic collisions with randomly located atoms. 
The corresponding angular distribution was calculated by Moli\`{e}re \cite{Moliere}, 
while the ionization energy loss distribution, by Landau \cite{Landau}, with subsequent refinements summarized in \cite{Bethe,Fano-AnnRev,Scott,Mott-Massey,Leroy-Rancoita,Sigmund-V2,PDG2020}.
However, those theories only correspond to simplest experimental configurations, in which the fast particle emerging from the target is directed into a single large detector measuring only one of the numbers characterizing the particle motion at the expense of erasing the rest of the information. 

%The outgoing high-energy particle flow after passage of a well-collimated beam through a not too thick target remains sufficiently narrow, and thus can normally be intercepted by no more than one detector.\footnote{In contrast to problems of particle stopping.}
%Simplest types of detectors measure either the charged particle deflection angle or its ionization energy loss. 

Modern thin detectors, easily penetrable by fast charged particles, 
sensitive even to small deflection angles and energy losses, 
and connectable into coincidence schemes, 
allow more detailed reconstruction of the particle motion history. 
For instance, a target made of a semiconductor material can serve as an \textit{in situ} detector of ionization energy losses \cite{Bak}, %(counted proportionally to the current of secondary electrons and holes)
while the particle incidence and emergence angles can be measured in the same event by a sequence of thin pixelized plates placed upstream and downstream the target (a ``telescope'' -- cf. \cite{Berger}). 
For such more advanced setups, Moli\`{e}re and Landau theories do not completely describe the particle distribution, 
insofar as they do not bring out possible correlation between the deflection angle and the energy loss. 
One should then evaluate the complete two-variable probability distribution function allowing for correlation.
The correlated function is also needed for some applied problems on fast charged particle passage through matter
%, including calculation of 3d spatial distribution of stopped particles 
(see, e.g., \cite{Remizovich-Rogozkin-Ryazanov,Kabachnik}). 

%and tracking devices, though, allow simultaneous measurement of position and energy loss (dEdx detectors [ ]). 
%When a few of such detector plates are included into a coincidence scheme (a telescope), by reconstructing the particle trajectory, one can  simultaneously measure its deflection angle and ionization loss in each plate. There are also setups, in which that is done naturally. 
%For instance, in experiments on channeling in long straight crystals, when the crystal is made of silicon, the channeled particles are often discriminated by their ionization losses in situ by a surface current detection . 
%It thus seems timely to generalize the theory to incorporate the correlation between the charged particle multiple scattering angles and ionization energy losses. 

Specific mechanisms of angle-energy loss correlation depend on the particle energy. 
Historically the first kind thereof, being due to the curvature extension of the particle path (``detour'') in a slab target, increasing with the increase of the emergence angle, was predicted in \cite{Pomeranchuk,Yang} (for a more detailed treatment, see \cite{Nakatsuka,Remizovich-Rogozkin-Ryazanov}). 
Since correlation of this kind is quadratic in the particle deflection angles, 
which are inversely proportional to the large particle momentum, it becomes negligible at high energy.
% it quickly diminishes with the increase of the latter. 

Another correlation mechanism is owing to sampling a higher electron density by particles passing 
%through regions 
near atomic nuclei, where the deflecting Coulomb field is stronger 
\cite{Meyer-Klein-Wedell,Jakas,Shindo,Remizovich-Ryazanov-Frolov}. 
It is pertinent to slow (with velocities $v\lesssim10^{-2}c$) ions, 
which act on the encountered atomic electrons semiclassically. 
Juxtaposing the single-collision mean energy loss $\Delta\epsilon(r_{\perp})$ as a function of the impact parameter $r_{\perp}$ with respect to the atomic nucleus \cite{Kabachnik}, 
and single-collision deflection angle $\chi(r_{\perp})$, 
one derives the elementary correlation between $\Delta\epsilon$ and $\chi$. 
%, assuming it to survive after rescatterings in the finite-thickness target 
With the increase of the incident ion energy, the semiclassical character of the scattering process eventually breaks down, but experimentally, the correlation was found to persist out to ion velocities $v\sim10^{-1}c$ \cite{Ishiwari,Sakamoto-1992-Cu}, where all the above mentioned mechanisms, as well as the target thickness nonuniformity \cite{Mertens-Krist},
%(a texture or roughness of the surface)
 seem to be not enough to explain the data \cite{Gras-Marti,Yamashita,Kabachnik}.

%Experiments with protons and ions on carbon [Hogberg,Beauchemin-Drouin,Ellmer-Sturm] and higher-$Z$ foils [Eckardt,Ishiwari,Sakamoto-1992-Cu,Fama2001], nonetheless, observed a positive correlation between the particle mean energy loss and the deflection angle even at energies when the ``detour'' correlation must already be extinct. This effect was then attributed to

%, but that hypothesis was put to doubt by computer simulation [Gras-Marti,Kabachnik], as well as the conjecture [Mertens-Krist] that the effect might be  merely due to a . 
%Full understanding of the origin of the effect thus remains challenging, particularly for energetic projectiles and low-$Z$ targets (for low energies and high $Z$, see []).
%Attempts to visually understand the origin of the correlation thus did not come to an immediate success, and computer simulation did not shed enough light on the issue.

At sufficiently high energies, when both scattering and ionization processes are quantum mechanical, 
it becomes possible to compare contributions from different correlation mechanisms by counting the minimal necessary number of interactions of the fast charged particle with electrons and the nucleus within a single atom in the medium. 
The mentioned correlation of the electron density with the location of the deflecting atomic nucleus requires at least two-photon exchange, 
in which one photon is exchanged with the static but highly charged nucleus, 
efficiently deflecting the incident charged particle, 
while another one is exchanged with the knocked-out electron, transferring it energy. 
But a positive correlation must arise even at the single-photon exchange level, when a virtual photon knocks out an atomic electron, imparting it transverse momentum and energy simultaneously. 
That mechanism, being of the lowest order in the small electromagnetic coupling constant, 
should dominate in the high-energy, quantum-mechanical domain.
%, similarly to the dominance of non-radiative processes over radiative ones

Besides comparing magnitudes of different effects in single scattering, it is important to take into account that targets for energetic, highly penetrating particles are usually made thick enough, in order to increase the deflection angle or the interaction rate. 
Certain types of angle-energy correlation may then be ruined by 
%For not too thin targets, most important here will be not even the strength of the coupling (because there will be 
multiple interactions with different atoms. 
In this regard, it is essential that Coulomb scattering of pointlike charged particles involves significant momentum transfers in single scatterings, 
contributing both to Rutherford ``tails'' at large deflection angles \cite{Moliere,Bethe,Scott} and to an anomalous (logarithmically modified) diffusion at typical angles \cite{Bond-Moliere}. 
That should support the dominance of single-photon exchange in the correlation.

In problems of high-energy multiple scattering in matter, it may thus important to investigate angle-energy loss correlation induced by a charged pointlike particle scattering on individual atomic electrons (hard incoherent scattering). 
This effect is expected to be the strongest for incident electrons or positrons, 
because a projectile of the same mass as the struck electrons will be able to transfer them a large fraction of its kinetic energy in ``head-on'' collisions. 
As for the target materials, more favourable should be low-$Z$ ones, for which the relative role of incoherent scattering increases. 
%A similar effect is known for heavy ions [Xia], when ... .

The objective of the present paper is to calculate the angle-energy loss correlation for relativistic electrons or positrons traversing low-$Z$ amorphous matter. 
This process is governed by a spatially uniform transport equation, solvable by conventional integral transformation techniques (Sec. \ref{sec:transp-eq}). 
In the multiple scattering regime, the solution further simplifies based on the same logarithmic approximations as in Moli\`{e}re, Fano, and Landau theories, whose ranges of validity have a broad enough intersection. 
It proves that for characterization of the substance, there is no need to introduce phenomenological parameters other than the elastic and inelastic screening angles $\chi_{a}$ and $\chi_{in}$, and the mean excitation energy $I_{\delta}$ including the density correction $\delta$. 
Furthermore, $I_{\delta}$ only contributes to an overall shift in the mean energy loss (just like for Landau distribution), whereas $\chi_{a}$ and $\chi_{in}$ coalesce into a single parameter (like in Fano theory). 
In Secs. \ref{sec:analysis}--\ref{sec:asymptotics}), we investigate the distribution function dependence on the two kinematic variables characterizing the final electron, and on two parameters ($Z$ and thickness) characterizing the target. 
We find that the correlation remains substantial even in thick targets, but its underlying mechanism differs from extension of the path or impact-parameter-mediated correlation discussed formerly.
We will also study the correlation in terms of conditional mean values for the two-variable distribution function (Sec. \ref{sec:mean-values}).
%, which in special experimental setups can be measured directly. 

%For self-consistent treatment of inelastic contribution to multiple Coulomb scattering, one should take into account that inelastic scattering the screening radius is different [Fano], but eventually they coalesce into a single parameter. 

%As for the dependence on $I$, it proves to be rather trivial, not affecting the distribution shape.
%Therefore, the joint angle-energy distribution depends on two parameters -- the traversed target thickness, measured in units of Moli\`{e}re angles (as in the Moli\`{e}re theory) and $Z$, which determines the fraction of inelastic scattering events, containing the correlation. 
%As is rather obvious, the smaller $Z$, the larger the correlation.

%\newpage

\section{Solution of the transport equation}\label{sec:transp-eq}

The correlated process of small-angle scattering of a fast electron on atoms of an amorphous substance, 
and the associated ionization energy loss, may be described by a transport equation
\begin{eqnarray}\label{transp-eq}
\frac{\partial}{\partial l}f(\bm\theta,\epsilon,l)=-n_a\int d\sigma_{el}(\chi)\left[f(\bm\theta,\epsilon,l)-f(\bm\theta-\bm\chi,\epsilon,l)\right]\nonumber\\
-n_a\iint d\sigma_{in}(\chi,\Delta\epsilon)\left[f(\bm\theta,\epsilon,l)-f(\bm\theta-\bm\chi,\epsilon-\Delta\epsilon,l)\right],\,\,\nonumber\\
\end{eqnarray}
where the first line represents contributions from elastic collisions, whereas the second line, from inelastic ones (atom excitation or knock-out of an atomic electron).
%In (\ref{transp-eq})
Here $f$ is the probability distribution function depending on the fast electron cumulative deflection angle $\bm\theta$ and cumulative ionization energy loss $\epsilon$, and normalized by 
\begin{equation}\label{normalization-epsilon-0infty}
\int d^2 \theta \int_0^{\infty}d\epsilon f=1.
\end{equation} 
$d\sigma_{el}$ and $d\sigma_{in}$ are the elastic and inelastic scattering differential cross sections depending on the single-scattering deflection angle $\bm\chi$ and energy transfer $\Delta\epsilon$. 
Also, $l$ designates the length of the electron path in the target, and $n_a$ the density of  atoms. 
It is presumed that $\theta\ll1$, and $\epsilon$ is much smaller than the fast particle energy, 
wherewith $d\sigma_{el}$ and $d\sigma_{in}$ virtually do not depend on the latter, 
as in Moli\`{e}re and Landau theories.

Besides that, fast electrons emit electromagnetic radiation, accompanying their scattering (bremsstrahlung). 
In principle, bremsstrahlung correlates with the electron scattering angle, as well \cite{Maximon-tagged-photons,Bond-anisotr-bremsstr}. 
But compared to elastic scattering, radiation emission probability is suppressed by a factor $e^2/\hbar c=1/137$. 
It is known that radiative energy losses dominate at high electron energies, because even a single photon can take away a large fraction of the electron energy. 
However, we will be interested in the energy loss distribution near the Landau peak (cf. \cite{Bak}), where radiative losses can be neglected \cite{Bak,Berger}.\footnote{Formally, $d\sigma_{in}$ in Eq. (\ref{transp-eq}) can include radiative processes (cf. \cite{Bond-multiphot}), 
but below we shall make specific assumptions about $d\sigma_{in}$, 
pertinent to the ionization process.}
Furthermore, soft radiation is suppressed by the density effect, 
whereas hard radiation freely escapes from the target-detector. 
In this respect, the situation is completely similar to that in Moli\`{e}re and Landau theories. 

%energy losses due to bremsstrahlung will be disregarded in the present treatment, since  and radiation reabsorption in the same target, at large photon energies, is even less probable. 
%It is essential, though, that , so, the target thickness has to be much smaller than the radiation length (for low-$Z$ amorphous materials amounting a few decimeters). 
%Ultimately, radiative and non-radiative energy losses can be unified within a common framework (see, e.e., \cite{Bond-multiphot} and refs. therein).

%It is characteristic of our approach that Eq. (\ref{transp-eq}) treats atomic electrons, as well as nuclei, as uniformly distributed, in spite that electrons in cold matter cluster around atomic nuclei. 
%The justification is that the incident electron, due to its small electric charge, interacts with not more than one constituent per atom -- either an electron or the nucleus. Then the impact parameter between this constituent and the incident electron is random, while the next atom is again encountered at a random impact parameter, etc. 
%At that, the screening within one atom is taken into account both in $d\sigma_{el}$ and $d\sigma_{in}$.

For our purposes, it is thus sufficient to deal with transport equation (\ref{transp-eq}), 
and complete it by an initial condition
\begin{equation}\label{}
f(\bm\theta,\epsilon,0)=\delta(\bm\theta)\delta(\epsilon),
\end{equation}
corresponding to a monokinetic initial beam. 
The solution of this problem is obtained as usual, by applying Fourier and Laplace transformations, 
which reduce (\ref{transp-eq}) to a first-order ordinary linear differential equation, 
solved by an exponential $e^{-l\kappa(\bm b, s )}$,
where\footnote{To reconcile nomenclatures used in Moli\`{e}re and Landau theories, and make them more naturally connected, we introduce here some new notations. 
%Our $b$ variable corresponds to $\eta$ in Moli\`{e}re's work , 
%while our $2mp^{-2}s$ was designated by $p$ in Landau's paper \cite{Landau}. 
%But $p$ is reserved herein for notation of the momentum, while $\eta$, on a par with $y$, for reduced $b^2$ (see below). 
%The rest of our notations comply with conventional ones. 
}
\begin{eqnarray}\label{kappa-def}
\kappa(\bm b, s )=n_a \int d\sigma_{el}(\chi) \left(1-e^{-i\bm{b}\cdot\bm{\chi}}\right)\qquad\qquad\qquad\quad\nonumber\\
+n_a \iint d\sigma_{in}(\chi,\Delta\epsilon) \left(1-e^{-i\bm{b}\cdot\bm{\chi}- 2mp^{-2} s \Delta\epsilon}\right)
\end{eqnarray}
($\bm b$- and $ s $-factorization). 
The inverse transformation yields a formal solution of our problem:
\begin{eqnarray}\label{double-Fourier}
\frac{p^2}{2m} f(\bm\theta,\epsilon,l)=\frac{1}{(2\pi)^2}\int d^2b\qquad\qquad\qquad\qquad\qquad\nonumber\\
\times \frac{1}{2\pi i}\int_{-i\infty}^{i\infty} d s  e^{i\bm{b}\cdot\bm{\theta}+2mp^{-2} s \epsilon}e^{-l\kappa(\bm b, s )}.
\end{eqnarray}
A factor $2mp^{-2}$ at $s\epsilon$ in the exponent, where $m$ and $p$ are the electron mass and momentum (which also degrades negligibly with the increase of $l$), has been introduced in Eqs. (\ref{kappa-def}), (\ref{double-Fourier}) for further convenience, in order to render $s$ dimensionless and commensurable with $b^2$. 
Granted that neither $d\sigma_{el}$ nor $d\sigma_{in}$ depend on the $\bm\chi$ azimuth, the azimuthal integration yields zero-order Bessel function of the product of $b=|\bm b|$ and  $\chi=|\bm \chi|$:
\begin{eqnarray}\label{fblambda=expJ0e}
\kappa(b, s )
=n_a\int d\sigma_{el}(\chi) \left[1-J_0(b\chi)\right]\qquad\qquad\qquad\nonumber\\
 +n_a\iint d\sigma_{in}(\chi,\Delta\epsilon) \left[1-J_0(b\chi)e^{-2mp^{-2}  s  \Delta\epsilon }\right].
\end{eqnarray}

The generic 2-variable dependence $d\sigma_{in}(\chi,\Delta\epsilon)$ takes account of all the effects, which can give rise to an angle-energy loss correlation. 
Which of them survives under conditions of multiple scattering will be seen next.

\paragraph{Multiple-scattering approximation}

In general, $d\sigma_{el}$ and $d\sigma_{in}$ are complicated functions of $\chi$ and $\Delta\epsilon$, reflecting the shell structure of atoms, and possibly even their binding effects in the solid. 
But under the conditions of multiple scattering, accumulated $\theta$ and $\epsilon$ are much greater than typical contributing $\chi$ and $\Delta\epsilon$ in the single-scattering cross sections. 
Given that  $b\sim \theta^{-1}$ and $s\sim \frac{p^2}{2m \epsilon }$, 
arguments $b\chi\sim \chi/\theta$ and $2mp^{-2} s \Delta\epsilon\sim \Delta\epsilon/\epsilon$ of functions  in (\ref{fblambda=expJ0e}) are typically small, 
so, differences $1-J_0(b\chi)$ and $1-J_0(b\chi)e^{- 2mp^{-2} s \Delta\epsilon}$ may be expanded in them to the leading orders:
\begin{equation}\label{1-J0=chi2}
1-J_0(b\chi)
\simeq \frac{b^2}{4}\chi^2,
\end{equation}
\begin{equation}\label{1-J0exp=chi2}
1-J_0(b\chi)e^{-2mp^{-2} s \Delta\epsilon}
\simeq \frac{b^2}{4}\chi^2+2mp^{-2} s \Delta\epsilon.
\end{equation}
However, straightforward application of such a procedure under conditions of Coulomb scattering leads to a logarithmic divergence of the integral, as long as both the elastic and inelastic differential scattering cross sections at large $\chi$ feature Rutherford asymptotics:\footnote{In treatment of electron scattering on atomic nuclei, we neglect nuclear size effects \cite{nucl-size} and spin corrections at wide scattering angles \cite{chimax}.
In contrast to the primary electron, knocked-out electrons in (\ref{dsigmainel-chic2}) are regarded as nonrelativistic, wherewith their dispersion law is $\Delta\epsilon=\frac{p^2\chi^2}{2m}$. As in the case of the Landau distribution, that is sufficient for description of the straggling far from the upper end of its spectrum. 
}
%\begin{equation}\label{Rutherford-asympt-el}
%d\sigma_{el}(\chi,\Delta\epsilon)\simeq \frac{8\pi Z^2 \alpha^2}{p^2} \frac{d\chi}{\chi^3},
%\end{equation}
%\begin{equation}\label{Rutherford-asympt}
%d\sigma_{in}(\chi,\Delta\epsilon)\simeq \frac{8\pi\alpha^2}{p^2} \frac{d\chi}{\chi^3}d\Delta\epsilon\delta\left(\Delta\epsilon-\frac{p^2}{2m}\chi^2\right).
%\end{equation}
%In product with $n_al$, $n_e l$, the pre-factors are conventionally written
\begin{equation}\label{dsigmael-chic2}
n_a ld\sigma_{el}(\chi)\underset{\chi\gg\chi_a}\simeq 
n_a ld\sigma_{Ruth}(\chi)
=2Z^2\bar\chi_c^2 \frac{d\chi}{\chi^3},
\end{equation}
and
\begin{equation}\label{dsigmainel-chic2}
n_ald\sigma_{in}(\chi,\Delta\epsilon)
\underset{\chi\gg\chi_a}\simeq 2Z \bar\chi_c^2
\delta\left(\Delta\epsilon-\frac{p^2}{2m}\chi^2\right)
\frac{d\chi}{\chi^3}d\Delta\epsilon.
\end{equation}
Here $\frac{d\sigma_{Ruth}}{d\chi}=\frac{8\pi Z^2e^4}{p^2v^2\chi^3}$ is the Rutherford cross section for scattering on a bare nucleus 
with charge $Ze$, and $v$ the particle velocity. 
Factor \cite{Fano-PR}
\begin{equation}\label{bar-chic2-def}
\bar\chi_c^2=\frac{4\pi e^4 n_a l }{p^2v^2}%=\frac{\chi_c^2}{Z(Z+1)},
\end{equation}
common both for elastic and inelastic processes, 
includes all the dependencies on the properties of the target except $Z$. 
Note that $Z$-dependence of the elastic scattering contribution (\ref{dsigmael-chic2}) is quadratic, 
reflecting its coherent nature, whereas inelastic scattering contribution (\ref{dsigmainel-chic2}), 
which is incoherent, is proportional to the number $Z$ of electrons per atom. 
Direct implementation of asymptotics (\ref{dsigmainel-chic2}) is the key property of our approach. %, as compared to approaches proposed for ions, where the energy loss is treated classically, as a single-valued and bounded function of impact parameter with respect to the atomic nucleus \cite{Meyer-Klein-Wedell,Jakas,Yamashita,Shindo,Remizovich-Ryazanov-Frolov}.
%Another common notation is [Bethe]

%accumulates all the thickness, density and energy dependencies, while $Z$-dependence can not be absorbed to an overall factor simultaneously in (\ref{dsigmael-chic2}) and (\ref{dsigmainel-chic2}). 

To treat the logarithmically divergent integrals accurately, one needs to reach 
%Taking into account the $\chi^{-1}$ scaling of the integrands obtained when factors (\ref{dsigmael-chic2}), (\ref{dsigmainel-chic2}) along with (\ref{1-J0=chi2}), (\ref{1-J0exp=chi2}) are inserted into  (\ref{fblambda=expJ0e}), the corresponding integrals have to be evaluated within 
the next-to-leading logarithmic accuracy (NLLA).
The standard procedure is to break the semi-infinite $\chi$ integration interval by an intermediate $\chi_1$, such that $\chi_{a},\chi_{in}\ll \chi_1\ll \bar\chi_c$, and  treat the large-$\chi$ contribution exactly, not resorting to expansions (\ref{1-J0=chi2}), (\ref{1-J0exp=chi2}), 
whereas the small-$\chi$ contribution -- phenomenologically [beyond Rutherford asymptotics (\ref{dsigmael-chic2}), (\ref{dsigmainel-chic2})].

\paragraph{Soft scattering contribution}
In the inner region $\chi<\chi_1$, expansions (\ref{1-J0=chi2}), (\ref{1-J0exp=chi2}) are safe to use. 
Additive constants in logarithmic $\chi_1$-dependencies of the integrals can be related to phenomenological parameters via definition
\begin{equation}\label{}
d\sigma_{el}=\mathsf{q}(Z,p\chi/\hbar)d\sigma_{Ruth}, \quad \mathsf{q}|_{\chi\sim\chi_1}=1, \quad \mathsf{q}(Z,0)=0
\end{equation}
for the elastic scattering function $\mathsf{q}$, further definition
\begin{equation}\label{}
\int_0^{\chi_1}\mathsf{q}\left(Z,\frac{p\chi}{\hbar}\right)\frac{d\chi}{\chi}=\ln\frac{\chi_1}{\chi'_{a}}+\gamma_{\text{E}}-1
\end{equation}
for the Moli\`{e}re's screening angle $\chi'_{a}(Z)$ \cite{Moliere,Bethe},\footnote{Following the notation of \cite{Bethe}, the prime at $\chi'_{a}$, discriminates it from the unprimed value $\chi_a=\chi'_a e^{1/2-\gamma_{\text{E}}}$, whose advantage is that it coincides with the inverse screening radius times $p^{-1}$ for Born scattering in a purely exponentially screened Coulomb potential \cite{Moliere}. 
But since such a potential is rather artificial in atomic physics, not being realized even for hydrogen atom, we work from the outset with the primed notations, in which the final results have simpler form. 
A similar relationship is implied for $\chi'_{in}$.}  
similar definition
\begin{equation}\label{}
\int_0^{\infty}d\Delta\epsilon
\frac{d\sigma_{in}}{d\Delta\epsilon d\chi}=\frac{d\sigma_{in}}{d\chi}=\mathsf{S} \left(Z,\frac{p\chi}{\hbar}\right)\frac{1}{Z} \frac{d\sigma_{Ruth}}{d\chi},
\end{equation}
\[
\mathsf{S}|_{\chi\sim\chi_1}=1, \quad \mathsf{S}(Z,0)=0
\]
for the inelastic scattering function $\mathsf{S}$, further  definition
\begin{equation}\label{}
\int_0^{\chi_1}\mathsf{S}\left(Z,\frac{p\chi}{\hbar}\right) \frac{d\chi}{\chi}=\ln\frac{\chi_1}{\chi'_{in}}+\gamma_{\text{E}}-1
\end{equation}
for the corresponding Fano angle $\chi'_{in}(Z)$ \cite{Fano-PR},\footnote{Assuming electrons to be sufficiently fast, and targets to have low or moderate $Z$, 
here we will neglect so-called Coulomb corrections, i.e., the cross section nonlinear dependence on $Ze^2/\hbar v$. 
Then, constants $\chi_{a}$ and $\chi_{in}$ are determined solely by atomic formfactors. 
For their evaluation at low $Z$, Thomas-Fermi model is too crude, and Hartree-Fock calculations are necessary \cite{Tollestrup-Monroe}. 
For hydrogen, one analytically evaluates $\ln {\chi_{in}}/{\chi_{a}}=\int_0^{\infty}\frac{d\chi}{\chi}(\mathsf{q}-\mathsf{S})=-5/6$, wherewith $\chi_{in}=e^{-5/6}\chi_{a}\approx0.4\chi_{a}$, 
${p\chi_{at}}/{\hbar}={2}/{ea_B}$, 
\begin{equation}\label{pchi'at-H}
\frac{p\chi'_{at}}{\hbar}=\frac{p\chi_{at}}{\hbar}e^{\gamma_{\text{E}}-1/2}=\frac{2}{a_B}e^{\gamma_{\text{E}}-3/2}
\end{equation}
 [with $\chi'_{at}$ defined by Eq. (\ref{chiat-def}) below].
} 
and
\begin{equation}\label{ioniz-potential-def}
\iint_{\chi<\chi_1}d\sigma_{in}\Delta\epsilon
=\frac{4\pi Z e^4}{mv^2}\left(\ln\frac{p\gamma\chi_1}{I_\delta}-\frac{v^2}{2c^2}\right),  
\end{equation}
\begin{equation}\label{lnIdelta-def}
\ln I_\delta=\ln I+\frac12\delta
\end{equation}
for the mean excitation energy $I(Z)$ (see \cite{Sauer-Sabin-Oddershede} and refs. therein). 
Density correction $\delta(Z,\gamma)$ in (\ref{lnIdelta-def}) accounts for dispersive dielectric susceptibility of the medium \cite{Leroy-Rancoita}.\footnote{At $\gamma\to\infty$, the behavior of $\delta(Z,\gamma)$ is such that
\[
\ln\frac{p}{I_\delta}=\ln\frac{m}{\hbar\omega_p(Z)}
+\frac12,
\]
where $\omega_p=\sqrt{4\pi Ze^2 n_a/m}$ is the plasma frequency of the medium, so, $\delta/2$ and $\ln\gamma$ compensate each other, and the logarithmic growth in Eq. (\ref{ioniz-potential-def}) saturates.} 
Combined, that gives
\begin{eqnarray}\label{el-sum-b-lambda}
n_a l\int_0^{\chi_1}d\chi \frac{d\sigma_{el}}{d\chi} \left[1-J_0(b\chi)\right]
\simeq
n_a l\frac{b^2}{4}\int_0^{\chi_1}d\chi \frac{d\sigma_{el}}{d\chi}\chi^2\nonumber\\
=\frac{Z^2\bar\chi_c^2b^2}{2} \left(\ln\frac{\chi_1}{\chi'_{a}}+\gamma_{\text{E}}-1\right),\qquad
\end{eqnarray}
\begin{eqnarray}\label{inel-sum-b-lambda}
n_a l \iint_{\chi<\chi_1} d\sigma_{in}\left[1-J_0(b\chi)e^{-2mp^{-2} s \Delta\epsilon}\right]\qquad\qquad\qquad\nonumber\\
\simeq
n_a l\frac{b^2}{4}\iint_{\chi<\chi_1} d\sigma_{in}\chi^2
+2mp^{-2} n_a l  s \iint_{\chi<\chi_1} d\sigma_{in} \Delta\epsilon\quad\nonumber\\
=2Z\bar\chi_c^2 \left[\frac{b^2}{4}\left(\ln\frac{\chi_1}{\chi'_{in}}+\gamma_{\text{E}}-1\right)
+ s  \left(\ln\frac{p\gamma\chi_1}{I_\delta}-\frac{v^2}{2c^2}\right)\right]\! .\nonumber\\
\end{eqnarray}

Notably, this part does not give rise to any angle-energy loss correlation, including that mentioned in the Introduction (sampled higher electron density in regions of stronger deflecting field near atomic nuclei). 
The reason is that this contribution is concentrated at limited energy and momentum transfers, but the lowest-order expansion in their Fourier-reciprocal variables involves no cross terms. 
If the diffusion was normal, that would be the end of the story for thick targets. 
%mean that under conditions of multiple scattering no correlation survives.

\paragraph{Hard scattering contribution}
There exists, however, an equally important contribution from the hard scattering region $\chi>\chi_1$. 
For its treatment, it is justified to apply Rutherford asymptotics (\ref{dsigmael-chic2}), (\ref{dsigmainel-chic2}), but the integrals 
\begin{equation}\label{chi1intinfty-dsigmain}
n_a l\int_{\chi_1}^\infty d\sigma_{el}\left[1-J_0(b\chi)\right]\simeq 2Z^2\bar\chi_c^2
\int_{\chi_1}^{\infty} \frac{d\chi}{\chi^3} \left[1-J_0(b\chi)\right],
\end{equation}
\begin{eqnarray}\label{integral-app}
n_a l\iint_{\chi>\chi_1}d\sigma_{in}\left[1-J_0(b\chi)e^{- s \chi^2}\right]\qquad\qquad \nonumber\\
\simeq 
2Z\bar\chi_c^2\int_{\chi_1}^{\infty}\frac{d\chi}{\chi^3}\left[1-J_0(b\chi)e^{- s \chi^2}\right]
\end{eqnarray}
need to be evaluated without resorting to Bessel and exponential function expansions (\ref{1-J0=chi2}), (\ref{1-J0exp=chi2}).

Integral (\ref{chi1intinfty-dsigmain}) is exactly the same as that arising in the Bethe-Moli\`{e}re theory, with the known result:
\begin{equation}\label{eq15}
\int_{\chi_1}^{\infty} \frac{d\chi}{\chi^3} \left[1-J_0(b\chi)\right]%\nonumber\\
\underset{b\chi_1\to0}\simeq
\frac{b^2}{4}\left(\ln \frac{2}{b\chi_1}+1-\gamma_{\text{E}}\right).
\end{equation}
Integral (\ref{integral-app}) is somewhat more sophisticated, and is evaluated in Appendix \ref{app:sum}:
\begin{eqnarray}\label{appendix-finalresult}
\int_{\chi_1}^{\infty} \frac{d\chi}{\chi^3} \left[1-J_0(b\chi)e^{- s \chi^2}\right]%\qquad\qquad\qquad\qquad\quad\nonumber\\
\underset{b\chi_1, s \chi_1^2\to0}\simeq
\frac{b^2}{4}+\frac{ s }{2}e^{-{b^2}/{4 s }}\nonumber\\
-\frac12
\left( s +\frac{b^2}{4}\right)\left[\ln  s \chi_1^2+\gamma_{\text{E}}
+\text{Ein}\left(\frac{b^2}{4 s }\right)\right],\quad
%\nonumber\\
%\equiv
%\left( s +\frac{b^2}{4}\right)\left[\ln \frac{2}{b\chi_1}
%- \frac12E_1\left(\frac{b^2}{4 s }\right)+\frac12-\gamma_{\text{E}}\right]\quad\nonumber\\
%-\frac{ s }{2}\left(1-e^{-{b^2}/{4 s }}\right)+\frac{b^2}{8},
\end{eqnarray}
with $\text{Ein}$ the complementary exponential integral function specified by Eq. (\ref{Ein-def}).
Again, the $\chi_1$-dependence here is simple logarithmic. 
As for $b$ and $ s $ dependencies, in contrast to Eqs. (\ref{el-sum-b-lambda}), (\ref{inel-sum-b-lambda}), they get intermixed. 
It is this mixing that induces a correlation between $\theta$ and $\epsilon$.

\paragraph{The combined result}
Ultimately, piecing together (\ref{chi1intinfty-dsigmain}), (\ref{eq15}) with (\ref{el-sum-b-lambda}), and (\ref{integral-app}), (\ref{appendix-finalresult}) with (\ref{inel-sum-b-lambda}), we cancel the delimiting parameter $\chi_1$, and are left with
%\begin{eqnarray*}\label{}
%l\kappa(b, s )= Z\bar\chi_c^2 \Bigg\{\frac{b^2}{2}\left[Z\ln\frac{2}{\chi'_{a}b}+\ln\frac{2}{\chi'_{in}b}-\frac12 E_1\left(\frac{b^2}{4 s }\right)\right]\nonumber\\
%+ s \left[2\ln\frac{2p\gamma}{bI_{\delta}}-1-E_1\left(\frac{b^2}{4 s }\right)-2\gamma_{\text{E}}+e^{-\frac{b^2}{4 s }}\right]\Bigg\}\nonumber\\
%= Z\bar\chi_c^2 \Bigg\{\frac{b^2}{2}\left[Z\ln\frac{2}{\chi'_{a}b}+\ln\frac{1}{\chi'_{in}}-\frac12 \text{Ein}\left(\frac{b^2}{4 s }\right)-\frac12\ln s +\frac12\gamma_{\text{E}}\right]\nonumber\\
%+ s \left[2\ln\frac{p\gamma}{I_{\delta}}-1
%-\text{Ein}\left(\frac{b^2}{4 s }\right)+\ln\frac{1}{ s }
%-\gamma_{\text{E}}+e^{-\frac{b^2}{4 s }}\right]\Bigg\}.
%\end{eqnarray*}
\begin{eqnarray}\label{}
l\kappa(b, s )%=\frac{Z^2\bar\chi_c^2b^2}{2} \left(\ln \frac{2}{b\chi_{a}}+\frac12-\gamma_{\text{E}}\right)\nonumber\\
%+2Z\bar\chi_c^2 \Bigg\{\frac{b^2}{4}\left(\ln\frac{1}{\chi_{in}}-\frac12\right)
%+ s  \left(\ln\frac{p\gamma}{I_\delta}-\frac{v^2}{2}\right)\nonumber\\
%+\frac{b^2}{4}+\frac{ s }{2}e^{-\frac{b^2}{4 s }}
%-\frac12
%\left( s +\frac{b^2}{4}\right)\left[\ln  s +\gamma_{\text{E}}
%+\text{Ein}\left(\frac{b^2}{4 s }\right)\right]\Bigg\}\nonumber\\
=Z\bar\chi_c^2 \Bigg\{\frac{b^2}{2}\left(Z \ln \frac{2}{b\chi'_{a}}
+\ln\frac{1}{\chi'_{in}}+\gamma_{\text{E}}\right)\qquad\nonumber\\
+ s  \left(2\ln\frac{p\gamma}{I_\delta}+e^{-{b^2}/{4 s }}-\frac{v^2}{c^2}\right)\qquad\nonumber\\
-\left( s +\frac{b^2}{4}\right)\left[\ln  s +\gamma_{\text{E}}
+\text{Ein}\left(\frac{b^2}{4 s}\right)\right]\Bigg\}.\quad
\end{eqnarray}

The final integral can be cast in a more compact form by 
%a few transformations. 
%First, as in the pure multiple scattering theory, the sum of logarithms of screening angles can be combined into a single empirical constant $\chi'_{a}$ [Fano]
%\begin{equation}\label{chi'a-def}
%Z\ln\frac{1}{\chi'_{a}}+\ln\frac{1}{\chi'_{in}}=(Z+1)\ln\frac{1}{\chi'_{a}}.
%\end{equation}
%Then,
%\begin{eqnarray*}\label{}
%l\kappa(b, s )=\frac{\chi_c^2 }{Z+1}\Bigg\{\frac{b^2}{2}\left[(Z+1)\ln\frac{2}{\chi'_{a}b}-\frac12 E_1\left(\frac{b^2}{4 s }\right)\right]\nonumber\\
%+ s \left[2\ln\frac{2p}{bI}-E_1\left(\frac{b^2}{4 s }\right)-2\gamma_{\text{E}}+e^{-\frac{b^2}{4 s }}\right]\Bigg\}.
%\end{eqnarray*}
%It should be noted that $p\chi'_a\gg I\sim10Z$ eV. But change of $I$ leads only to a shift in $\epsilon$.
%It will be still expedient, though, to treat the two terms on the right-hand side of (\ref{chi'a-def}) separately, including the terms proportional to $Z$ to the elastic contribution.
changing $b$ and $s$ to reduced integration variables 
\begin{equation}\label{b,lambda-def}
y=Z^2 \bar\chi_c^2 b^2, \qquad  u=Z\bar\chi_c^2 s,
\end{equation}
which absorb the energy dependence and partly the thickness dependence. 
Accordingly, kinematic variable $\theta$ is changed to
\begin{equation}\label{Theta-def}
\Theta=\frac{\theta}{Z\bar\chi_c},
\end{equation}
and $\epsilon$ to 
\begin{eqnarray}\label{u-energyloss}
\lambda(Z,l,\epsilon)=\frac{2m}{p^2Z\bar\chi_c^2}\epsilon-\lambda_0,\label{lambda-def}%\\
%&\equiv&\frac{2\epsilon}{\alpha\gamma^2 \omega_p^2l}-\ln\frac{\alpha m\gamma^2\omega_p l}{I_{\delta}^2}+\gamma_{\text{E}}-\gamma^{-2},\nonumber
\end{eqnarray}
with a shift 
\begin{equation}\label{lambda0-def}
\lambda_0=\ln\frac{p^2\gamma^2 Z\bar\chi_c^2}{I_{\delta}^2}-\gamma_{\text{E}}+\gamma^{-2}
\end{equation}
arising after combining the term $\epsilon u$ in the exponent with the other terms linear in $u$.
%\begin{eqnarray*}\label{}
%l\kappa(y,u)=\frac{y}{4Z}\left[(Z+1)\ln\frac{4Z \chi_c^2}{(Z+1)\chi'^2_{a}y}- E_1\left(\frac{y}{4Zs}\right)\right]\qquad \nonumber\\
%+s\left[\ln\frac{1}{s}+\ln\frac{p^2\chi_c^2}{(Z+1)I^2}-\gamma_{\text{E}}-\text{Ein}\left(\frac{y}{4Zs}\right)+e^{-\frac{y}{4Zs}}\right].
%\end{eqnarray*}
%\begin{eqnarray*}\label{}
%l\kappa(b, s )= \frac{y}{2Z}\left(\frac{Z}{2} \ln \frac{4Z^2 \bar\chi_c^2}{y}
%+(Z+1)\ln\frac{1}{\chi'_a}+\gamma_{\text{E}}\right)\nonumber\\
%+s \left(2\ln\frac{p\gamma}{I_\delta}+e^{-\frac{y}{4Zs}}-v^2\right)\nonumber\\
%-\left(s+\frac{y}{4Z}\right)\left[\ln \frac{s}{Z\bar\chi_c^2}+\gamma_{\text{E}}
%+\text{Ein}\left(\frac{y}{4Zs}\right)\right]\nonumber\\
%=\frac{y}{4}\ln\frac{y_0}{y}
%+s \left(2\ln\frac{p\gamma\bar\chi_c}{I_\delta}+\ln Z-\gamma_{\text{E}}
%+\gamma^{-2}+e^{-\frac{y}{4Zs}}-1\right)\nonumber\\
%-\left(s+\frac{y}{4Z}\right)\left[\ln u+\text{Ein}\left(\frac{y}{4Zs}\right)\right].
%\end{eqnarray*}
In this notation, the distribution function expresses as
\begin{eqnarray}\label{f=intdy-intds}
\frac{p^2}{2m}Z^3 \bar\chi_c^4  f(Z,l,\theta,\epsilon)
=F(Z,y_0,\Theta,\lambda)
\qquad\qquad\quad\nonumber\\
=\frac{1}{4\pi}\int_0^{y_{\max}} dyJ_0(\sqrt{y}\Theta) e^{\Omega_{el}(y_0,y)}\nonumber\\
\times
\frac{1}{2\pi i}\int_{-i\infty}^{i\infty} du e^{\lambda u +\Omega_{in}(y/4Z,u)},
\end{eqnarray}
where dimensionless parameter  
\begin{equation}\label{y0-def}
y_0(l,Z)=4Z \left(\frac{ Z\bar\chi_c^2}{\chi'^2_{at}}\right)^{1+1/Z}e^{\gamma_{\text{E}}/Z}\gg 1,
\end{equation}
serves as a measure of the target thickness for given $Z$, and is independent of $v$, 
because in the ratio $\bar\chi_c/\chi'_{at}$ both $\bar\chi_c$ and the effective screening angle
\begin{equation}\label{chiat-def}
\chi'_{at}= \chi'^{\frac{Z}{Z+1}}_{a} \chi'^{\frac{1}{Z+1}}_{in}
\end{equation}
of elastic and inelastic scattering are reciprocal to $v$. 
Functions 
%\begin{subequations}
\begin{equation}\label{Sigmael-def}
\Omega_{el}(y_0,y)=-\frac{y}{4}\ln\frac{y_0}{y},
\end{equation}
%\begin{eqnarray}
%\Delta(Y,u)=-Y\left[\ln Y+\gamma_{\text{E}}
%+E_1\left(Y/u\right)\right]\qquad\qquad\qquad\nonumber\\
%-s\left[E_1\left(Y/u\right)+\ln\frac{Y}{s}+\gamma_{\text{E}}+1-e^{-Y/u}\right]\qquad\qquad\label{Delta-a}\\
%\equiv 
%-Y\left[\ln u+\text{Ein}\left(Y/u\right)\right]
%-s\left[\text{Ein}\left(Y/u\right)+1-e^{-Y/u}\right].
%\label{Delta-b}
%\end{eqnarray}
%\end{subequations}
\begin{equation}\label{Sigmain-def}
\Omega_{in}(Y,u)%=s\ln u-\Delta(Y,u)
=\left(u+Y\right)\left[\ln u+\text{Ein}\left(Y/u\right)\right]+u\left(1-e^{-Y/u}\right).
\end{equation}
are defined so that they vanish at the origin: 
\[
\Omega_{el}(y_0,0)=\Omega_{in}(0,0)=0.
\] 

As can be verified with the use of identities
\begin{equation}\label{intinftydlambda}
\int_{-\infty}^{\infty}d\lambda \frac{1}{2\pi i}\int_{-i\infty}^{i\infty} du e^{\lambda u+\Omega_{in}(Y,u)}=e^{\Omega_{in}(Y,0)},
\end{equation}
\begin{equation}\label{int0inftydThetaThetaint0inftydy}
\int_0^{\infty}d\Theta\Theta\int_0^{\infty}dy J_0(\sqrt{y}\Theta)e^{\Omega_{el}+\Omega_{in}}=2 e^{\Omega_{el}+\Omega_{in}}\Big|_{y=0}
\end{equation}
(being particular cases of inverse Fourier and Fourier-Bessel transformations),
$F$ is normalized to unity in variables $\Theta$ and $\lambda$:
\begin{equation}\label{F-normalization}
2\pi\int_0^{\infty}d\Theta\Theta\int_{-\infty}^{\infty}d\lambda F(Z,y_0,\Theta,\lambda)=1,
\end{equation}
in accordance with its interpretation as a probability density. 
Note that compared with (\ref{normalization-epsilon-0infty}), 
by virtue of a very rapid convergence of the $\lambda$-integral, 
its actual lower integration limit $-\lambda_0$, 
where $\lambda_0$ is logarithmically large, 
was replaced by $-\infty$ (as in Landau theory).

Before proceeding, let us analyze the structure of the obtained integral (\ref{f=intdy-intds}).
Even though one of its entries, $\Omega_{in}$, looks somewhat cumbersome, involving a special function $\text{Ein}$, its second $u$-derivative is very simple:
\begin{equation}\label{d2s=exp}
\frac{\partial^2}{\partial u^2}\Omega_{in}(Y,u)=\frac{1}{u}e^{-Y/u}.
\end{equation}
[This relation can be derived directly by double differentiating (\ref{appendix-finalresult}) over $s$, and setting thereupon $\chi_1=0$, 
granted that the integrand vanishes at this endpoint.]
We will take advantage of property (\ref{d2s=exp}) below. 
%It is positive, hence, the integral over the imaginary $u$ axis converges.

Next, it is worth noting that besides the dependence of the exponent $\Omega_{el}+\Omega_{in}$ on variables $y$ and $u$, 
in terms of which the transport equation has been factored, it depends yet on parameters $Z$ and $y_0$. 
In total, that amounts to a 4-variable dependence, but in Eq. (\ref{f=intdy-intds}) it actually splits into a sum of two functions, each depending only on two variables. 
That by no means implies any factorization of the resulting Fourier integral, 
because both $\Omega$'s depend on $y$. 
After the integration over $y$, the dependencies on $\Theta$ and $\lambda$ intermix. 
In this regard, it is worth reminding that $y$ is a rescaled square of $\bm b$ [Eq. (\ref{b,lambda-def})], 
where $\bm b$ may be thought of as an analogue of an impact parameter. 
The correlation between $\Theta$ and $\lambda$ may thus be attributed to correlations of each of them with $\bm b$, 
resembling impact-parameter-mediated correlation for slow ions (cf. Introduction). 
It should be understood, however, that $\bm{b}$ is not a physical transverse coordinate $\bm{r}_{\perp}$, 
but a Fourier reciprocal to $\bm\theta$ at the probability level.

The dependence of $\Omega_{in}$ on $y$ and $Z$ only through their ratio reflects the fact that aggregate scattering on atomic electrons, 
because of its incoherent character, is $Z$ times weaker than that on atomic nuclei. 
(Recall that a coherent scattering factor $Z^2$ has been included in the definition of $y$, 
while the incoherent scattering factor $Z$ in the definition of $u$.) 
Besides that, $\Omega_{in}$ involves functions depending only on a ratio $Y=y/4Zu$.
As we will see in Sec. \ref{sec:central-region}, this implies asymptotic dependence only on the ratio of $\lambda$ and $Z$.
%The dependence of the integrand on a ratio of two variables generally implies some scaling properties, which after the integration over $y$, survive only approximately, transferring on variable $\lambda$.

\begin{figure}
\includegraphics{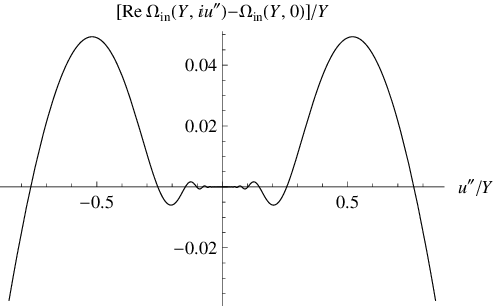}
\caption{\label{fig:Sigmain} The real part of the exponent in the $u$-integral (\ref{f=intdy-intds}), 
with $\Omega_{in}$ defined by Eq. (\ref{Sigmain-def}), for $Y=y/4Z$ and imaginary $u=iu''$. 
%Continuation to the negative $u''$ domain is symmetrical.
}
\end{figure}

To determine whether Eq. (\ref{f=intdy-intds}) is suitable for numerical evaluation, it is also necessary to assess the rate of convergence of its $y$- and $u$-integrals. 
The $u$-dependence of the integrand is determined by asymptotics of the real part of $\Omega_{in}$, 
which is defined by Eq. (\ref{Sigmain-def}). 
Therein, $\text{Ein}(Y/u)\underset{|u|\to\infty}\to 0$, $1-e^{-Y/u}\underset{|u|\to\infty}\to 0$, 
hence, at imaginary integration variable $u=iu''$ and large $|u''|$, $\Omega_{in}$ behaves as $\mathfrak{Re}(u\ln u)=-\frac{\pi}{2}|u''|$, linearly tending to $-\infty$. 
Hence, at infinity the $u$-integral converges exponentially, which is rapid enough. 
In the opposite limit of small imaginary $u$, functions $e^{-Y/u}$ and $\text{Ein}(Y/u)$ rapidly oscillate, and even though the exponent at that does not blow up (see Fig. \ref{fig:Sigmain}), 
it may be expedient to introduce a small lower cutoff, e.g., $|u''|_{\min}\sim 10^{-5}$. 

\begin{figure}
\includegraphics{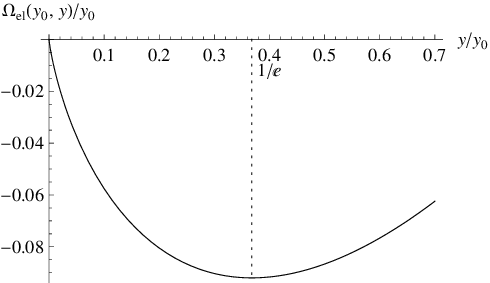}
\caption{\label{fig:Sigmael} The behavior of the exponent in the $y$-integral (\ref{f=intdy-intds}), 
with $\Omega_{el}(y_0,y)$ defined by Eq. (\ref{Sigmael-def}). 
Dotted line indicates the location of its minimum. 
The corresponding $y$ sets the upper scale for $y_{\max}$ [Eq. (\ref{ymaxlesssimy0})].
%Continuation to the negative $u''$ domain is symmetrical.
}
\end{figure}

As for the $y$-integral in Eq. (\ref{f=intdy-intds}), it is primarily determined by the behavior of $\Omega_{el}$. 
In the limit $y\to0$, both $\Omega_{el}$ and $\Omega_{in}$ are finite, but $\Omega_{el}$ begins to grow indefinitely when $y>y_0/e$ (see Fig. \ref{fig:Sigmael}), threatening with divergence of the entire integral. 
Such a divergence was absent in the original integral, 
so, it is a byproduct of the small-$y$ approximation. 
%The mathematical meaning of $y_0$ is that it sets for , hence the subscript $0$.
Just like in the Moli\`{e}re theory, 
it can be precluded merely by introducing a sizable upper cutoff $y_{\max}$, being not larger than $\sim y_0/e$:
\begin{equation}\label{ymaxlesssimy0}
1\ll y_{\max}\lesssim y_0/e.
\end{equation} 
The result of the integration is then insensitive to $y_{\max}$. 
The subscript 0 actually indicates that $y_0$ serves as the upper limit for $y$ integration.
Taking into account the inequality $\Omega_{el}>-0.08\, y_0$ (see Fig. \ref{fig:Sigmael}), 
the necessary condition for the target to be physically thick ($-\min\Omega_{el}\gg 1$) is \cite{Scott}
\begin{equation}\label{y0gtrsim102}
y_0\gtrsim 10^2.
\end{equation}

As a practical guide, the correspondence between the target thickness and values of $y_0$ used in further examples is given Table \ref{table:thetainf}, for $Z\leq6$.

\begin{center}
\begin{table}[h]
\caption{\label{table:thetainf} Target thicknesses corresponding to different values of $y_0$, for low $Z$. 
The calculations are performed in the Born approximation (neglecting Coulomb corrections), 
based on atomic elastic scattering formfactors and inelastic scattering functions tabulated in \cite{Wang95}.}
%\footnotesize\rm
\centering
%\begin{ruledtabular}
\begin{tabular}{|c|c|c|c|c|}%was {@{}*{8}{l}}
\hline
$Z$ & ${p\chi'_{at}}/{\hbar}$ & $y_0=10^2$ & $y_0=10^4$ & $y_0=10^6$ \\
\hline
1  & 1.50 $\text{\AA}^{-1}$ & $l=0.28\,\frac{\text{mg}}{\text{cm}^2}$ &  $l=2.82\,\frac{\text{mg}}{\text{cm}^2}$ &  $l=28.2\,\frac{\text{mg}}{\text{cm}^2}$ \\
2  & 2.91 $\text{\AA}^{-1}$ & $l=2.29\,\frac{\text{mg}}{\text{cm}^2}$ &  $l=49.3\,\frac{\text{mg}}{\text{cm}^2}$ &  $l=1.06\,\frac{\text{g}}{\text{cm}^2}$ \\
3  & 3.10 $\text{\AA}^{-1}$ & $l=51.5\,\mu$m & $l=1.63$ mm & $l=5.15$ cm \\
4  & 3.44 $\text{\AA}^{-1}$ & $l=15.6\,\mu$m & $l=0.62$ mm & $l=2.47$ cm \\
5  & 3.75 $\text{\AA}^{-1}$ & $l=11.3\,\mu$m & $l=0.526$ mm & $l=2.44$ cm \\
6  & 3.88 $\text{\AA}^{-1}$ & $l=12.63\,\mu$m & $l=0.651$ mm & $l=3.37$ cm \\
%$7$  & 4.45618 &  &  &  \\
%$8$  & 4.78896 &  &  &  \\
\hline
\end{tabular}
%\end{ruledtabular}
\end{table}
\end{center}

\section{Distribution shapes}\label{sec:analysis}

\begin{figure}
\includegraphics{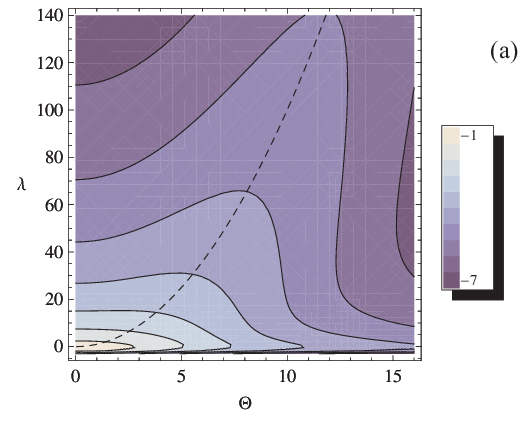}
\includegraphics{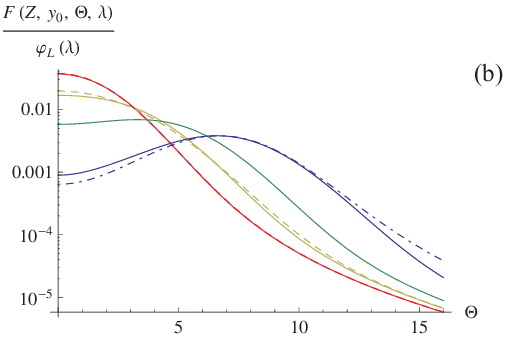}
\includegraphics{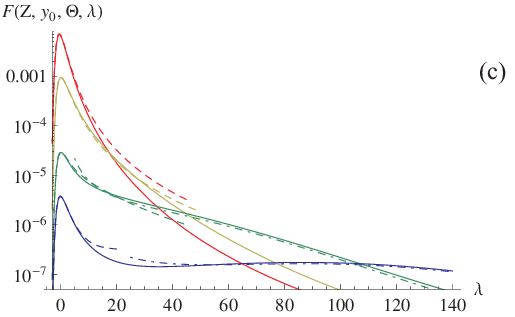}
\caption{\label{fig:PlotZ1} 
Correlated angle-energy loss distribution of electrons multiple scattered in hydrogen ($Z=1$) target with thickness parameter $y_0=10^4$ (see Table \ref{table:thetainf}). 
a). Contour plot of $\log F$, with $F$ given by Eq. (\ref{f=intdy-intds}). 
Dashed parabola, the midline of the spur [Eq. (\ref{lambda=ZTheta2})].
b). Angular distributions at fixed values of the ionization energy loss straggling variable $\lambda$. Solid curves, calculation by exact formula (\ref{f=intdy-intds}), at $\lambda=0$ (red), $\lambda=7$ (yellow), $\lambda=20$ (green), $\lambda=50$ (blue). 
Dashed, approximation (\ref{quasi-factorization}) for same values of $\lambda$. 
Dot-dashed, approximation (\ref{Fproptog}), (\ref{betaM}). 
c). Ionization energy loss distribution at fixed values of the scattering angle. 
Solid curves, $\Theta=0$ (red), $\Theta=4$ (yellow), $\Theta=8$ (green), $\Theta=16$ (blue). 
Dashed and dot-dashed curves correspond to the same approximations as in (b).
}
\end{figure}

\begin{figure}
\includegraphics{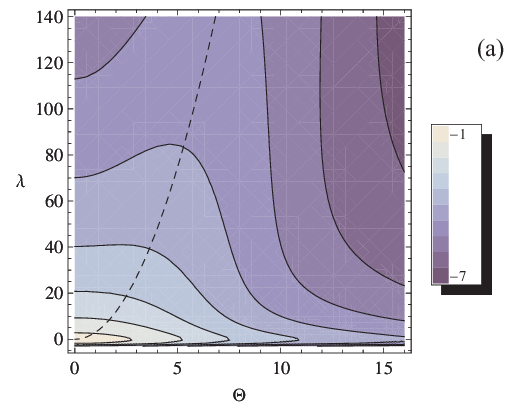}
\includegraphics{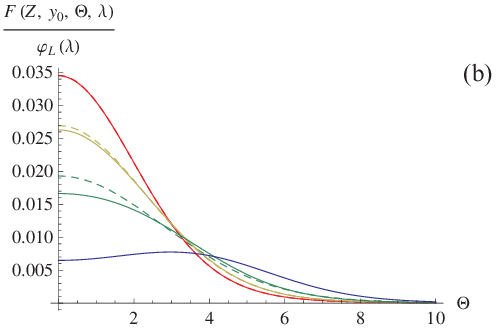}
\includegraphics{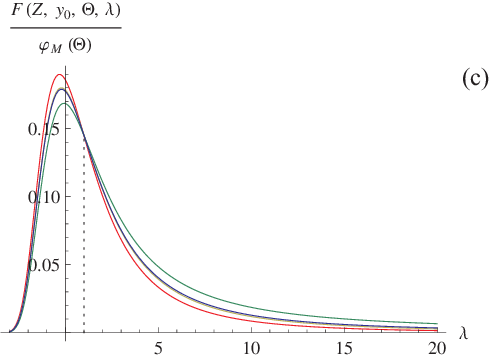}
\caption{\label{fig:PlotZ3} 
Correlated angle-energy loss distribution of  electrons multiple scattered in lithium ($Z=3$) target with thickness parameter $y_0=10^4$. 
a). Contour plot of $\log F$, with $F$ given by Eq. (\ref{f=intdy-intds}). 
Dashed parabola, the midline of the spur [Eq. (\ref{lambda=ZTheta2})].
b). Angular distributions at fixed values of the straggling variable $\lambda$. Solid curves, calculation by exact formula (\ref{f=intdy-intds}), at $\lambda=0$ (red), $\lambda=7$ (yellow), $\lambda=20$ (green), $\lambda=50$ (blue). 
Dashed, approximation (\ref{quasi-factorization}). 
%Dot-dashed, approximation (\ref{Fproptog}), (\ref{azim-aver-varphiM}). 
c). Ionization energy loss distribution at fixed values of the scattering angle. 
Solid curves, $\Theta=0$ (red), $\Theta=3$ (yellow), $\Theta=6$ (green), $\Theta=15$ (blue).
In the moderate $\lambda$ domain, the latter is the closest to Landau distribution. 
All the straggling curves intersect at $\lambda=1$ (dotted vertical line). 
}
\end{figure}

\begin{figure}
\includegraphics{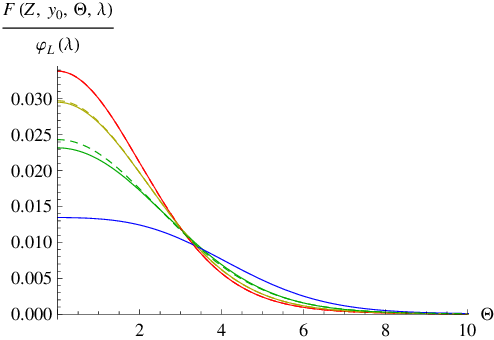}
\caption{\label{fig:PlotZ6} 
%a). Contour plot of $\log_{10} F$, with $F$ given by Eq. (\ref{f=intdy-intds}). b). 
Angular distributions of electrons multiple scattered in carbon ($Z=6$) target with thickness parameter  $y_0=10^4$, at fixed values of the ionization energy loss straggling variable $\lambda$: 
$\lambda=0$ (red), $\lambda=7$ (yellow), $\lambda=20$ (green), $\lambda=50$ (blue). Solid curves are calculated by exact formula (\ref{f=intdy-intds}). 
Dashed curves, approximation (\ref{quasi-factorization}) for same values of $\lambda$.  
%Dot-dashed, approximation (\ref{Fproptog}), (\ref{azim-aver-varphiM}). 
%c). Ionization energy loss distribution at a fixed scattering angle. Solid curve, $\Theta=0$ (red), $\Theta=3$ (yellow), $\Theta=6$ (green), $\Theta=9$ (blue).
}
\end{figure}

\begin{figure}
\includegraphics{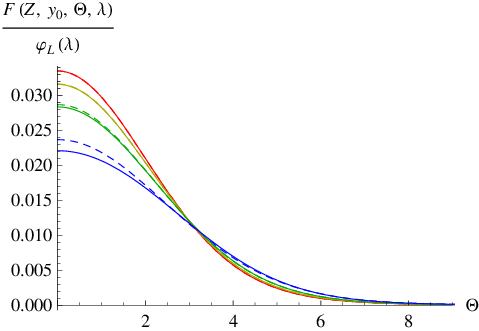}
\caption{\label{fig:PlotZ14} 
The same as Fig. \ref{fig:PlotZ6}, for silicon ($Z=14$) target with thickness parameter $y_0=10^4$. 
}
\end{figure}

%Rapid convergence of the double integral (\ref{f=intdy-intds}) makes its numerical sufficiently simple. 
Examples of density plots for the computed correlated distribution in the $\Theta$-$\lambda$ plane are presented in Figs. \ref{fig:PlotZ1}a, \ref{fig:PlotZ3}a. 
Their characteristic features are an elliptic concentration at moderate $\lambda,\Theta$ (formed in the process of multiple scattering), 
and a ``spur'' along the parabola 
\begin{equation}\label{lambda=ZTheta2}
\lambda=Z\Theta^2,
\end{equation}
extending to higher $\lambda$ and $\Theta$, 
and being due to single hard scattering. 
The oblique orientation of this spur is a vivid illustration that the joint distribution does not reduce to a product of single-variable, $\Theta$- and $\lambda$-dependent functions. 
One can also observe that when the spur merges with the dominant central spot, 
it somewhat distorts the latter diagonally. 
Further inspection of slices of those distributions along $\Theta$ and $\lambda$ axes (Figs. \ref{fig:PlotZ1}b,c, \ref{fig:PlotZ3}b,c, \ref{fig:PlotZ6}, \ref{fig:PlotZ14}) reveals that their shapes are qualitatively similar to Landau and Moli\`{e}re distributions (cf. Figs. \ref{fig:varphiL}, \ref{fig:varphiM} below), 
but the width of the conditional $\Theta$ distribution depends on $\lambda$, and vice versa. 
At sufficiently large $\lambda$, 
the maximum of the conditional $\Theta$-distribution is at $\Theta\neq 0$
(the slice of the spur -- see Figs. \ref{fig:PlotZ1}b,c, \ref{fig:PlotZ3}b).

The presence of two dissimilar concentration regions in the probability distribution hints that 
in each of them the double Fourier integral representation (\ref{f=intdy-intds}) 
may admit further simplifications under suitable approximations.
But in any case, the description of $F$ can not be simpler than that of its single-variable ($\lambda$- or $\Theta$-) projections, which are Landau and Moli\`{e}re distributions, known to be irreducible to elementary functions. 
The intrinsic simplicity of the latter owes instead to their stability -- invariance with respect to convolutions, as exemplified by Eq. (\ref{convol-phiL}) below. 
By virtue of that, Landau and Moli\`{e}re distributions can in fact serve as building blocks even in the two-variable case. 
Below we shall remind their definitions and basic properties.

\subsection{Landau distribution}

\begin{figure}
\includegraphics{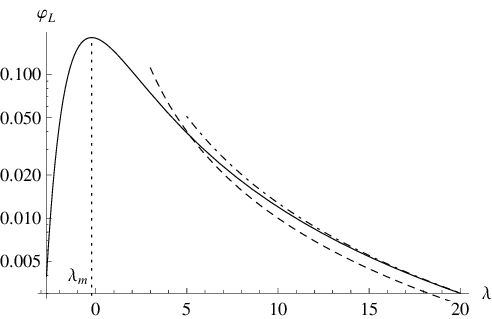}
\caption{\label{fig:varphiL} Log plot of Landau distribution (\ref{phiL-def}) (solid curve). 
Dashed curve, Rutherford asymptotics (\ref{varphiL-Ruth}) (approaching the solid curve at very large $\lambda$).  Dot-dashed, Rutherford asymptotics with power correction (\ref{varphiL-Ruth-NLO}), 
approaching the solid curve more rapidly. 
The most probable energy loss [Eq. (\ref{lambdam})], 
corresponding to the maximum of $\varphi_L$, is marked by the dotted vertical line.
}
\end{figure}

Integration of Eq. (\ref{f=intdy-intds}) over the full $\bm\Theta$ plane can be performed with the aid of  identity (\ref{int0inftydThetaThetaint0inftydy}), and gives
\begin{eqnarray}\label{phiL-def}
2\pi\int_0^{\infty}d\Theta\Theta F(Z,y_0,\Theta,\lambda)\qquad\qquad\quad\nonumber\\
=\frac{1}{2\pi i}\int_{-i\infty}^{i\infty} du e^{\lambda u +\Omega_{in}(0,u)}=\varphi_L(\lambda).
\end{eqnarray}
Here 
\begin{equation}\label{Sigmain0u}
\Omega_{in}(0,u)=u\ln u,
\end{equation}
whereby $\varphi_L$ is recognized to be Landau distribution \cite{Landau,Borsch-Supan}. It is normalized by
\begin{equation}\label{}
\int_{-\infty}^{\infty}d\lambda \varphi_L(\lambda)=e^{\Omega_{in}(0,0)}=1,
\end{equation}
has the group property
\begin{equation}\label{convol-phiL}
\int_{-\infty}^{\infty}d\lambda_1 \varphi_L(\lambda_1)\varphi_L(\lambda-\lambda_1)
=\frac12 \varphi_L\left(\frac{\lambda}{2}-\ln 2\right),
\end{equation}
and in the large energy loss region has the single-scattering, Rutherford symptotics\footnote{Asymptotics (\ref{varphiL-Ruth}) becomes sufficiently accurate only rather remotely (see Fig.~\ref{fig:varphiL}, dashed curve). Higher accuracy may be achieved by taking into account a correction term:
\begin{eqnarray}\label{varphiL-Ruth-NLO}
\varphi_L(\lambda)\underset{\lambda\to\infty}
\simeq \frac{1}{2\pi i}\int_{-i\infty}^{i\infty} du e^{\lambda u}\left(u\ln u+\frac12 u^2\ln^2 u\right)\nonumber\\
=\frac{1}{\lambda^2}
+\frac{2\ln \lambda+2\gamma_{\text{E}}-3}{\lambda^3}\qquad\qquad\qquad\quad\,\,\,\,
\end{eqnarray}
(displayed in Fig.~\ref{fig:varphiL} by the dot-dashed curve).
In the present paper, for simplicity, we will restrict our analysis to the leading-order asymptote (\ref{varphiL-Ruth}) and its counterparts for the joint distribution function (see Sec. \ref{subsec:large-u}). 
It should be minded that at moderately large $\lambda$, their accuracy is limited.
}
\begin{equation}\label{varphiL-Ruth}
\varphi_L(\lambda)\underset{\lambda\to\infty}\simeq \frac{1}{2\pi i}\int_{-i\infty}^{i\infty} du e^{\lambda u}u\ln u=\frac{1}{\lambda^2}.
\end{equation}

At $\lambda\to-\infty$, function $\varphi_L(\lambda)$ drops off faster than exponentially, as can be demonstrated by integration in the saddle-point approximation \cite{Landau}. 
The entire shape of $\varphi_L(\lambda)$ is markedly asymmetric (see Fig. \ref{fig:varphiL}). 
Its maximum is achieved at 
\begin{equation}\label{lambdam}
\lambda=\lambda_m\approx-0.22
\end{equation}
(the most probable energy loss). In this point,
\begin{equation}\label{varphi''L(lambdam)}
\varphi_L(\lambda_m)\approx0.18,\quad \varphi'_L(\lambda_m)=0,\quad \varphi''_L(\lambda_m)\approx-0.079.
\end{equation}

\subsection{Moli\`{e}re distribution}\label{subsec:Moliere}

Similarly, integration over $\lambda$ in Eq. (\ref{f=intdy-intds}) proceeds with the use of identity (\ref{intinftydlambda}), yielding
\begin{eqnarray}\label{}
\int_{-\infty}^{\infty}d\lambda F(Z,y_0,\Theta,\lambda)
=\frac{1}{4\pi}\int_0^{y_{\max}} dyJ_0(\sqrt{y}\Theta)\qquad \nonumber\\
\times e^{\Omega_{el}(y_0,y)+\Omega_{in}\left(y/4Z,0\right)}.\quad
\end{eqnarray}
Since $\Omega_{in}\left(y/4Z,0\right)=\frac{y}{4Z}\left(\ln \frac{y}{4Z}+\gamma_{\text{E}}\right)$ has the same structure as $\Omega_{el}(y_0,y)$ given by Eq. (\ref{Sigmael-def}), all those terms in the exponent can be amalgamated by collecting $\ln y+\text{const}$ factors of $-\frac{y}{4}$ as
\begin{equation}\label{collect-lny+const}
\ln\frac{y_0}{y}-\frac{1}{Z}\left(\ln\frac{y}{4Z}+\gamma_{\text{E}}\right)%\qquad\qquad\qquad\nonumber\\
%=\frac{Z+1}{Z}\ln\frac{1}{y}+\ln y_0+\frac{1}{Z}\ln 4Z-\frac{\gamma_{\text{E}}}{Z}\nonumber\\
=\frac{Z+1}{Z}\ln\frac{4Z^2\bar\chi_c^2}{\chi'^2_{at} y}, %\qquad\qquad\qquad\quad\,\,\,\,\,
\end{equation}
where definition (\ref{y0-def}) of $y_0$ was used.
Then, it is expedient to change the integration variable from $y$ to
\begin{equation}\label{eta-def}
\frac{Z+1}{Z}y=\eta,
\end{equation}
whereafter the pure angular distribution expresses as
\begin{equation}\label{intduF}
\int_{-\infty}^{\infty}d\lambda F(Z,y_0,\Theta,\lambda)
=\frac{Z}{Z+1}\varphi_M\left(\eta_0,\Psi\right),
\end{equation}
where 
\begin{equation}\label{Psi-def}
\Psi=\frac{\theta}{\chi_c}=\sqrt{\frac{Z}{Z+1}}\Theta
\end{equation}
is another version of the reduced scattering angle, 
and
\begin{equation}\label{eta0-def}
\eta_0=\frac{4\chi_c^2}{\chi'^2_{at}}
\end{equation}
is the reduced thickness. 
In definitions (\ref{Psi-def}), (\ref{eta0-def}) 
\begin{equation}\label{chic2-def}
\chi_c^2(Z,l,n_a,p)=Z(Z+1)\bar\chi_c^2(l,n_a,p) \equiv \frac{4\pi Z(Z+1) e^4 n_a l }{p^2v^2}
\end{equation}
is the conventional parameter of Moli\`{e}re theory \cite{Bethe}, differing from $\bar\chi_c^2$ by a $Z(Z+1)$ factor. 
Function
\begin{equation}\label{Moliere-def}
\varphi_M(\eta_0,\Psi)=\frac{1}{4\pi}\int_0^{\eta_{\max}} d\eta J_0\left(\sqrt{\eta}\Psi\right) e^{-\frac{\eta}{4}\ln\frac{\eta_0}{\eta}}
\end{equation}
(with $\eta_{\max}\sim y_{\max}$, the sensitivity to which is weak) is Moli\`{e}re distribution normalized by
\begin{equation}\label{}
2\pi\int_0^{\infty}d\Psi\Psi \varphi_M(\eta_0,\Psi)=1.
\end{equation}
%All the target characteristics $Z$ and $l$ enter here via a single parameter $\eta_0$, which somewhat differs from $y_0$, and depends on $l$ linearly. Similarly, variable $\Psi$ somewhat differs from $\Theta$ defined by Eq. (\ref{Theta-def}). 
It appears that variables $y_0$ and $\Theta$ are the best suited for the two-variable case,
whereas $\eta_0$ and $\Psi$ are natural for the $\lambda$-integrated or large-$\lambda$ distributions (see below). 

\begin{figure}
\includegraphics{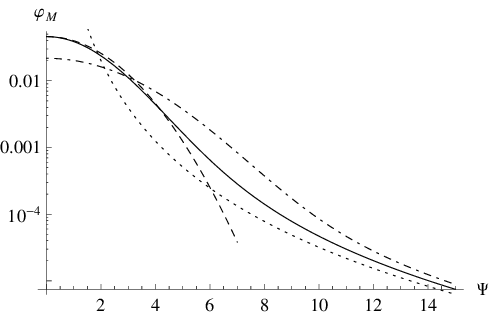}
\caption{\label{fig:varphiM} Log plot of Moli\`{e}re distribution 
(\ref{Moliere-def}) for $\eta_0=10^3$ (solid curve). 
Dashed curve, its leading logarithmic approximation (\ref{Moliere-LLA}). Dot-dashed, Moli\`{e}re distribution for $\eta_0=10^6$. 
Dotted curve, Rutherford asymptotics (\ref{varphi_M-Rutherford}), valid for any $\eta_0$.
}
\end{figure}

For understanding the behavior of $\varphi_M$, it is rewarding to note that it obeys an ordinary diffusion equation, with $\ln \eta_0$ playing the role of ``time'':
\begin{equation}\label{diffusion-eq-varphiM}
\frac{\partial \varphi_M}{\partial \ln \eta_0}=\frac{1}{4}\Delta_{\Psi}\varphi_M , \qquad \Delta_{\Psi} =\frac{\partial^2 }{\partial \Psi^2}+\frac{1}{\Psi}\frac{\partial}{\partial \Psi}.
\end{equation}
This equation only describes a hard part of the angular diffusion, whereas soft, normal diffusion enters via the explicit square root dependence of $\Psi$ on $l$ [see Eqs. (\ref{Psi-def}), (\ref{chic2-def})]. 
In spite of having only a logarithmic thickness dependence, hard diffusion in a thick target is as significant as the soft one.
Yet despite being governed by a normal diffusion equation, $\varphi_M$ is non-Gaussian in $\Psi$, insofar as it does not correspond to a $\delta(\bm{\Psi})$ initial condition. 
In particular, for all $\eta_0$ it exhibits the same large-$\Psi$ Rutherford ``tail" 
\begin{equation}\label{varphi_M-Rutherford}
\varphi_M(\eta_0,\Psi)\underset{\Psi\to\infty}\simeq \frac{1}{16\pi}\int_0^{\infty} d\eta\eta\ln \eta J_0\left(\sqrt{\eta}\Psi\right) = \frac{1}{\pi \Psi^4}
\end{equation}
(see Fig. \ref{fig:varphiM}, dotted curve). 
It is illegitimate to evolve $\varphi_M$ back to small $\ln \eta_0$, anyway, because $\eta_0$ needs to be large to ensure negligible dependence of $\varphi_M$ on the upper integration limit in definition (\ref{Moliere-def}). 
Therewith, hard diffusion does not reduce to a rescaling of deflection angle with time.

For very thick targets, $\eta_0\to\infty$, the simplest approximation for $\varphi_M$ is leading logarithmic (dating back to Williams \cite{Williams,Mott-Massey}), when in the exponential entering Eq.~(\ref{Moliere-def}) $\ln\eta$ is neglected compared to $\ln\eta_0$:
\begin{eqnarray}\label{Moliere-LLA}
\varphi_M(\eta_0,\Psi)\underset{\eta_0\to\infty}
\simeq \frac{1}{4\pi}\int_0^{\infty} d\eta J_0\left(\sqrt{\eta}\Psi\right) e^{-\frac{\ln\eta_0}{4}\eta} \nonumber\\
= \frac{1}{\pi \ln\eta_0} e^{-\Psi^2/\ln\eta_0}.\qquad\qquad\quad\,\,\,
\end{eqnarray}
The accuracy of this Gaussian approximation is illustrated in Fig. \ref{fig:varphiM} by the dashed curve. 
It can be satisfactory for description of low-statistics experiments, 
when only the central angular region is visible, but as we shall see below, is generally too crude for treatment of angle-energy loss correlation effects. 

More accurate approximations can be obtained by expanding in Eq. (\ref{Moliere-def}) the exponential $e^{\frac{\eta}{4}\ln\eta}$ in powers of $\frac{\eta}{4}\ln\eta$ and integrating termwise (an equivalent of Moli\`{e}re expansion \cite{Moliere,Bethe,Scott}). 
In what follows, we will need to know the value of $\pi\varphi_M$ at $\Psi=0$:
\begin{eqnarray}\label{piMolierePsi0-NLLA}
\pi\varphi_M(\eta_0,0)%= \frac{1}{4}\int_0^{\infty} d\eta  e^{-\frac{\eta}{4}(\ln\eta_0-\ln\eta)} 
%\qquad\qquad\quad\,\,\,\nonumber\\
\underset{\ln\eta_0\gg1}\simeq 
\frac{1}{4}\int_0^{\infty} d\eta  e^{-\frac{\eta}{4}\ln\eta_0}\left(1+\frac{\eta}{4}\ln\eta\right)\qquad\nonumber\\
=\frac{1}{\ln\eta_0}+\frac{1}{\ln^2\eta_0}\left(\ln\frac{4}{\ln\eta_0}+1-\gamma_{\text{E}}\right),\quad
\end{eqnarray}
or its inverse
\begin{equation}\label{1piMolierePsi0-NLLA}
\frac{1}{\pi\varphi_M(\eta_0,0)}\underset{\ln\eta_0\gg1}\simeq 
\ln\eta_0+\ln\frac{\ln\eta_0}{4}-1+\gamma_{\text{E}}.
\end{equation}
Approximation (\ref{1piMolierePsi0-NLLA}) works well already at $\eta_0\gtrsim 10^2$ corresponding to condition (\ref{y0gtrsim102}), 
whereas the leading logarithmic approximation [the first term of (\ref{1piMolierePsi0-NLLA})] can become insufficiently accurate at $\eta_0\gtrsim 10^4$, 
due to the lack of the growing term $\ln\ln\eta_0$.

\subsection{Normalized correlation function}\label{subsec:g-1}

Single-variable distributions $\varphi_L$, $\varphi_M$ quoted above can be used to formally define the angle-energy loss correlation. 
According to (\ref{phiL-def}) and (\ref{intduF}), the correlation between those variables would be absent if the distribution function factorized into a product of the corresponding single-variable distributions:
\begin{equation}
F(Z,y_0,\Theta,\lambda)\approx 
\frac{Z}{Z+1} \varphi_L(\lambda)\varphi_M\left(\eta_0,\Psi\right),
\end{equation}
with
\[
\Psi=\sqrt{\frac{Z}{Z+1}}\Theta,
\quad 
\eta_0=\frac{Z+1}{Z}\left(4Z e^{-\gamma_{\text{E}}}\right)^{\frac{1}{Z+1}}y_0^{\frac{Z}{Z+1}}\sim y_0.
\]
The relative deviation from this structure, i.e., deviation of the normalized correlation function
\begin{equation}\label{would-be-factoriz}
 g(Z,y_0,\Theta,\lambda)=\frac{Z+1}{Z}\frac{F(Z,y_0,\Theta,\lambda)}{\varphi_L(\lambda)\varphi_M\left(\eta_0,\Psi\right)}%\qquad\qquad\quad\nonumber\\
%\equiv\frac{Z+1}{Z}\frac{F(Z,y_0,\Theta,\lambda)}{\varphi_L(\lambda)\varphi_M\!\left[\frac{Z+1}{Z}\left(4Z e^{-\gamma_{\text{E}}}\right)^{\frac{1}{Z+1}}y_0^{\frac{Z}{Z+1}},\sqrt{\frac{Z}{Z+1}}\Theta\right]}\nonumber\\
\end{equation}
from unity thus provides a measure of the correlation.

\begin{figure}
\includegraphics{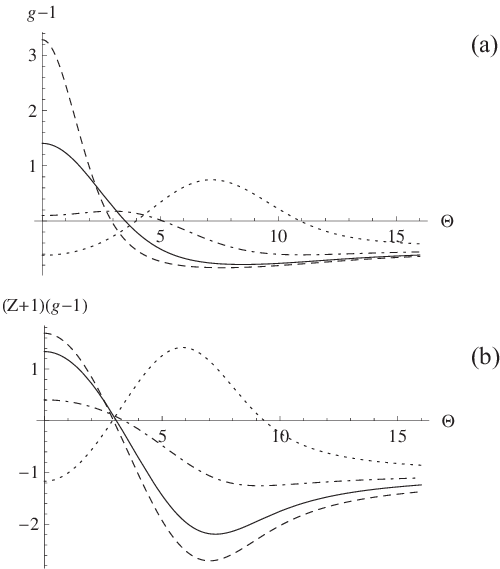}
\caption{\label{fig:g-1} (a). Deviation of the normalized correlation function (\ref{would-be-factoriz}) from unity, for $Z=1$ and $y_0=10^4$. 
Curves: $\lambda=-3$ (dashed), $\lambda=0$ (solid), $\lambda=7$ (dot-dashed), $\lambda=20$ (dotted).
(b). $(Z+1)(g-1)$ for $Z=14$ and the same values of $y_0$ and $\lambda$. 
}
\end{figure}

Inspection of Fig. \ref{fig:g-1}, where difference $g-1$ is plotted vs $\Theta$, shows that in general, it is nowhere negligible, unless $Z$ is very large. 
At high $\Theta$, it tends to a negative constant value
\begin{equation}\label{g-1asympt}
g-1\underset{\Theta\to\infty}\to-\frac{1}{Z+1},
\end{equation}
as will be proven in Sec. \ref{subsec:fixedlambda-largeTheta}.
At low $\Theta$, difference $g-1$ appears to be of the same order, but its sign depends on the value of $\lambda$ (see Figs. \ref{fig:g-1}a,b).  
The physical reason for the correlation to be of the order of $\frac{1}{Z+1}$ is that it is caused only by scattering on atomic electrons, whose aggregate contribution is $\sim Z$ times smaller than that from scattering on atomic nuclei.
It appears that product $(Z+1)(g-1)$ for $Z\geq6$ is nearly $Z$-independent for all $\Theta$, $\lambda$ (see Fig. \ref{fig:g-1}b).

It may thus be concluded that  the correlation between $\lambda$ and $\Theta$ is significant enough for light materials, including practical cases such as carbon ($Z=6$) and silicon ($Z=14$). 
At the same time, the results of this section indicate that there are two regions (the central one and the spur), 
in which the distribution function behaves quite differently. 
Those regions deserve an independent investigation, which will be provided in the next two sections.

\section{Moderate energy transfers. Large-thickness and large-$Z$ approximations}\label{sec:central-region}

%To gain more insight into the structure of the solution, we need to directly scrutinize the behavior of the double integral (\ref{f=intdy-intds}) in different regimes, corresponding to two main probability concentration regions manifest in Figs. \ref{fig:PlotZ1}a, \ref{fig:PlotZ3}a. 
%Each of those regions may be characterized by its own inherent approximations, in which $\varphi_M$ and $\varphi_L$ can  emerge in a natural way. 
Turning to investigation of the distribution function by regions, 
let us begin with the simpler case of the central region, 
where the bulk of the probability resides. 

This region (see  Figs. \ref{fig:PlotZ1}a, \ref{fig:PlotZ3}a) is populated primarily by events consisting of a sequence of many sufficiently probable soft scatterings in a thick target. 
At derivation of generic Eq. (\ref{f=intdy-intds}), 
it was already taken into account that under conditions of multiple scattering, typical contributing values of $b$ and $ s $ tend to zero with the increase of the target thickness. 
That has brought $\Omega_{el}$ to a small-$y$ limiting form, 
although typical values of the \emph{reduced} variable $y$ decrease only logarithmically. 
As for $\Omega_{in}$, it involves $\text{Ein}$ and an exponential function depending on the ratio $y/4Zu$. 
Being inversely proportional to $u$, which varies from $0$ to $\infty$, 
this ratio can be arbitrarily large or small. 
But in the presently considered central kinematic region, 
$\lambda$ is supposed to be limited (not too far from typical). 
Then, typical values of $u$ must be non-vanishing, and 
%Correspondingly (as was illustrated above, see Figs. \ref{fig:FTheta-fixedu}, \ref{fig:Fu-fixedTheta}), the bulk of the probability is concentrated at moderate reduced variables $\Theta$, $\lambda$. 
%Let us thus first investigate manifestations of the correlation in this region in more detail. That is possible by adopting further approximations admissible in this region. 
%In principle, with the increase of the target thickness, represented by parameter $y_0$, typical $y$ slowly tend to zero. 
%As for typical $u$, for not too large $\lambda$, they are not small. 
%Generally, they are also bounded from above ($u\lesssim1$, except for large negative $\lambda$ values, where the probability drops off very steeply), insofar as the $u$-integral convergence is furnished by the $\Omega_{in}(0,u)=u\ln u$ term in the exponent (see the end of Sec. \ref{sec:transp-eq}). 
at typical $y\underset{y_0\to\infty}\to0$, contributing values of arguments  $y/4Zu$ of functions entering $\Omega_{in}$ tend to zero. 
In order to retain the correlation, we develop $\Omega_{in}$ in this ratio to the next-to-leading, 1st order. 
Utilizing (\ref{Sigmain0u}) and the value of the derivative
\begin{equation}\label{}
\frac{\partial\Omega_{in}}{\partial Y}\bigg|_{Y=0}=2+\ln u,
\end{equation}
we expand
\begin{equation}\label{Sigmain-NLO-y}
\Omega_{in}(y/4Z,u) \underset{y/4Zu\ll 1}\simeq u\ln u+\frac{y}{4Z}\left(2+\ln u\right).
\end{equation}
Grouping (\ref{Sigmain-NLO-y}) with the rest of the terms in the exponent,
\[
\Omega_{el}+\lambda u+\Omega_{in}
\simeq \left(-\frac{y}{4}\ln\frac{y_0}{y}+\frac{y}{2Z}\right)
+\lambda u+\left(u+\frac{y}{4Z}\right)\ln u,
\]
we observe that the exponent involves only one mixing term of a gratifyingly simple form $y\ln u$.

\subsection{$\lambda$-dependence of the $\Theta$-distribution}

By virtue also of the fact that the found mixing term enters with a small factor $1/4Z$, 
it can be eliminated approximately by applying a relatively small shift of one of the integration variables:
\begin{equation}\label{}
u=\tilde u-\frac{y}{4Z}.
\end{equation}
After the corresponding linearization of the term $\tilde u\ln u\simeq \tilde u\ln\tilde u-y/4Z$, variables $y$ and $\tilde u$ approximately separate:
\begin{equation}\label{Sigmael+us+Sigmain}
\Omega_{el}+\lambda u+\Omega_{in}%\qquad\qquad\quad\nonumber\\
%\simeq-\frac{y}{4}\ln\frac{y_0}{y}-\frac{y}{4Z}(u-1)+u\tilde u+\tilde u\ln \tilde u\nonumber\\
\simeq
-\frac{y}{4}\ln\frac{y_0K}{y}+\lambda\tilde u+\tilde u\ln \tilde u,%\qquad\qquad
\end{equation}
where we have introduced a notation
\begin{equation}\label{K-def}
K(Z,\lambda)=e^{\frac{\lambda-1}{Z}}.
\end{equation}
Substitution of (\ref{Sigmael+us+Sigmain}) to double integral (\ref{f=intdy-intds}) brings it to a quasifactorized form
\begin{equation}\label{quasi-factorization}
F(Z,y_0,\Theta,\lambda)
\approx \varphi_L(\lambda)\varphi_M\left[y_0K(Z,\lambda),\Theta\right],
\end{equation}
where $\lambda$-dependence, besides $\varphi_L$, 
enters to the parameter of $\varphi_M$. 
The argument of $\varphi_M$ here is $\Theta$, in contrast to $\Psi$ entering to the right-hand side of Eq. (\ref{intduF}) for the $\lambda$-integrated case.

The validity condition for approximation (\ref{quasi-factorization}) is $\lambda\sim u^{-1}\ll 4Z/y\sim Z\ln y_0$, i.e.,
\begin{equation}\label{lambdallZlny_0}
\lambda\ll Z\ln y_0.
\end{equation}
It is important that here in the right-hand side $\ln y_0\gg1$, thus, the exponent of (\ref{K-def}) needs not be small. Insofar as the right-hand side in (\ref{lambdallZlny_0}) involves a factor of $Z$, 
it may as well be regarded as the large-$Z$ limit. 
At $Z=1$, condition (\ref{lambdallZlny_0}) is fulfilled marginally, 
so, the accuracy of (\ref{quasi-factorization}) for $Z=1$ is poorer.

From comparison of solid and dashed curves in Figs. \ref{fig:PlotZ1}b,c, \ref{fig:PlotZ3}b (for $\ln y_0=\ln 10^4\approx 9$), 
it is evident that approximation (\ref{quasi-factorization}) can remain tenable even for $\lambda\sim Z\ln y_0$, 
but inevitably breaks down at $\lambda\gg Z\ln y_0$, when the maximum of the distribution function is no longer at $\Theta=0$. 
The scaling dependence on $\frac{\lambda-1}{Z}$, though, can hold under more general conditions, as will be demonstrated in Secs. \ref{subsec:condit-lambda} and \ref{subsec:6A}. 

Structure (\ref{quasi-factorization}) permits a simple physical interpretation.
In the central region, an increase of $\lambda$ leads to a broadening of the angular distribution, which is equivalent to an effective increase of the electron path length (multiplication of the length parameter $y_0$ by a $\lambda$-dependent factor $K$).\footnote{E.g., (\ref{quasi-factorization}) can be represented in terms of the action of a length dilation operator
\[
\varphi_M(y_0K,\Theta)=K^{\frac{\partial }{\partial \ln y_0}}\varphi_M(y_0,\Theta)
=e^{\Psi_{\lambda-1}^2\frac{\partial }{\partial \ln l}}\varphi_M(y_0,\Theta),
\]
with $\Psi_{\lambda-1}^2=\frac{\lambda-1}{Z+1}$.
} 
That is reminiscent of the ``detour'' mechanism mentioned in the Introduction, with the proviso that the trajectory of a high-energy particle remains nearly straight (there is no appreciable ``detour''). 
The effect, instead, owes to the increase of relative contribution of events with large energy transfers 
(and correspondingly, large momentum transfers)
with an increase of observed $\lambda$. 
The mathematical stability of the shape of the Moli\`{e}re distribution ensures that this shape is unaffected by an admixture of harder scattering, which is why $\varphi_M$ appears in Eq.~(\ref{quasi-factorization}). 
% is stable under conditions of multiple Coulomb scattering, any evolution, including that at  fixed $\lambda$, must tend to $\varphi_M$ with some parameter. 
%More specifically, a convolution of any narrow distribution with a stable broad one gives a stable distribution of the same kind with a slightly increased width. 
%More precisely, typical $\Delta\Theta$, corresponding to $\lambda$, are independent of $Z$, whereas nuclear ones are $\propto Z$. 
%It must be understood, however, that at high $Z$, the correlation manifests itself only at proportionally large $\lambda$, where $\varphi_L(\lambda)$ is already small.
Finally, the fact that in (\ref{quasi-factorization}) $K$ depends solely on ratio $\frac{\lambda-1}{Z}$ is explained by recalling that $\lambda$ and $\Theta$ are defined so that for the hard contribution $\lambda/Z \sim \Theta^2$ [see Eq. (\ref{lambda=ZTheta2})], 
whereas the exponential form of $K$ owes to $\ln y_0$ serving as a ``time'' variable for $\Theta$-diffusion 
[cf. Eq. (\ref{diffusion-eq-varphiM})].

%It also deserves attention that the quasi-factorized expression (\ref{quasi-factorization}) depends only on 3 rather than 4 variables.

\subsection{$\Theta$-dependence of the $\lambda$-distribution}\label{subsec:condit-lambda}

Formula (\ref{quasi-factorization}) may be used, conversely, to determine the influence of a prescribed value of $\Theta$ on the $\lambda$-distribution. 
In the point $\lambda=1$, as Eqs. (\ref{quasi-factorization}), (\ref{K-def}) indicate, ratio $F(Z,y_0,\Theta,1)/\varphi_M(y_0,\Theta) \approx\varphi_L(1)$ does not depend on $\Theta$, $Z$, or $y_0$ at all. 
Another virtually fixed point is $\lambda\approx-3$, being effectively the leftmost edge of the distribution (see Fig. \ref{fig:PlotZ3}c). 
In other points, $F(Z,y_0,\Theta,\lambda)/\varphi_M(y_0,\Theta)$ depends, besides $\lambda$, on other variables. 

\begin{figure}
\includegraphics{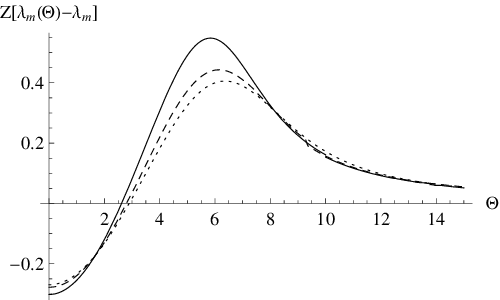}
\caption{\label{fig:MPEL-Theta} $\Theta$-dependence of the most probable energy loss [value of $\lambda$ at which $F(Z,y_0,\Theta,\lambda)$ reaches its maximum with respect to $\lambda$]. 
To reduce $Z$-dependence, the difference $\lambda_m(\Theta)-\lambda_m$ is multiplied by $Z$. 
Solid curve, for $Z=1$; dashed, $Z=3$; dotted, high-$Z$ approximation (\ref{lambdam-logder-varphiM}).
All the curves are built for $y_0=10^4$. 
}
\end{figure}

As Figs. \ref{fig:PlotZ1}c, \ref{fig:PlotZ3}c, and more clearly Fig. \ref{fig:MPEL-Theta} demonstrate, 
the increase of $\Theta$ leads to a \emph{nonmonotonic} variation of location of the point $\lambda_m(\Theta)$, in which $F(\Theta,\lambda)$ as a function of $\lambda$ reaches its maximum ($\Theta$-dependent most probable energy loss), i.e., obeys condition
\[
\frac{\partial}{\partial\lambda} F(Z,y_0,\Theta,\lambda)\Big|_{\lambda=\lambda_m(\Theta)}=0.
\]
In the large-$Z$ limit, this equation can be solved by expanding $\varphi_L$ around its maximum [see Eq. (\ref{varphi''L(lambdam)})]:
\[
\varphi_L(\lambda)\simeq \varphi_L(\lambda_m)\left[1+\frac{\varphi''_L(\lambda_m)}{\varphi_L(\lambda_m)}\frac{(\lambda-\lambda_m)^2}{2}\right],
\]
linearizing $\varphi_M$ in $\lambda$:
\[
\varphi_M(y_0K,\Theta)\simeq 
\varphi_M(y_0,\Theta)\left(1+\frac{\lambda-1}{Z}\frac{\partial\ln\varphi_M}{\partial\ln y_0}\right),
\]
and substituting this to Eq. (\ref{quasi-factorization}): 
\begin{eqnarray}\label{FphiM-expand}
F(Z,y_0,\Theta,\lambda)
\approx \varphi_L(\lambda_m)\varphi_M(y_0,\Theta)\qquad\qquad\qquad\qquad\nonumber\\
\times \left[1+\frac{\varphi''_L(\lambda_m)}{\varphi_L(\lambda_m)}\frac{(\lambda-\lambda_m)^2}{2}+\frac{\lambda-1}{Z}\frac{\partial\ln\varphi_M}{\partial\ln y_0}\right].\quad
\end{eqnarray}
Differentiating the right-hand side of (\ref{FphiM-expand}) by $\lambda$ and equating to zero, we find
\begin{equation}\label{lambdam-logder-varphiM}
\lambda_m(\Theta)\approx \lambda_m+\frac{1}{Z} \frac{\varphi_L(\lambda_m)}{\left|\varphi''_L(\lambda_m)\right|}
\frac{\partial\ln\varphi_M(y_0,\Theta)}{\partial\ln y_0},
\end{equation}
where ${\varphi_L(\lambda_m)}/{\left|\varphi''_L(\lambda_m)\right|}\approx 2.3$.

In physical terms, the nonmonotonic behavior of $\lambda_m(\Theta)$ may be interpreted as follows. 
At low $\Theta$, the contributing particle trajectories are straighter, wherewith typical partial momentum and energy transfer values are lower, implying that so must be the cumulative $\lambda_m(\Theta)$. 
At moderately large $\Theta$, in the semihard region \cite{Bond-Moliere}, there are several relatively large-angle scatterings, so, the particle trajectory is the most crooked, wherewith the energy loss is the highest. 
Finally, at $\Theta\to\infty$, when the trajectory is single-angle-shaped, $\lambda_m(\Theta)\to \lambda_m$, because even though the energy loss in this single wide-angle scattering event is large, in the straggling distribution it contributes beyond the region of typical losses. 

%\begin{figure}
%\includegraphics{\lambda-1ZscalingTheta=0}
%\includegraphics{\lambda-1ZscalingTheta=10}
%\caption{\label{fig:u-1Zscaling} Approximate scaling -- dependence on $\frac{\lambda-1}{Z}$. 
%Dotted curve, approximation (\ref{quasi-factorization}). 
%Dashed, $Z=1$.
%Solid, $Z=6$ (and higher).
%(a). $\Theta=0$; (b) $\Theta=10$; .
%}
%\end{figure}

\section{Hard scattering region. Large $\Theta$, $\lambda$}\label{sec:asymptotics}

Compared with the mild angle-energy loss correlation in the central region, described in the previous section, 
in the large-$\Theta$ and -$\lambda$ region the correlation is expected to tighten, 
due to the enhancement of the relative contribution of single hard scattering events, 
each being intrinsically highly correlated. 
But even there, multiple scattering effects do not disappear completely, 
because observation of a large-$\lambda$ event ascertains only that a single hard scattering on an atomic electron has occurred, 
whilst the number of soft elastic and inelastic scatterings may be arbitrary, and generally is large. 
Hence, even in the hard scattering region, effects of multiple scattering on the $\Theta$-$\lambda$ distribution can be substantial, 
and need be taken into account. 

To be self-consistent, again, we start with the double integral (\ref{f=intdy-intds}), 
and formally evaluate it in the corresponding limit. 
It should be pointed out that two possible ways of going to infinity in the $\Theta$-$\lambda$ plane, increasing one of the coordinates while holding the other fixed, do not commute. 
Indeed, limit $\Theta\to\infty$ implies typical $y\to0$ in the integrand, whereas limit $\lambda\to\infty$, typical $u\to0$. 
But since $\Omega_{in}$ defined by Eq. (\ref{Sigmain-def}) involves functions depending on the ratio $y/u$, the result must depend on the order, in which limits $y\to0$ and $\lambda\to0$ are taken. 
The noncommutability of limits $\Theta\to\infty$ and $\lambda\to\infty$ is also evident from Fig.~\ref{fig:PlotZ1}a, where at large $\lambda$ the spur separates two asymptotic voids. 
Let us thus consider both limiting sequences by turn.

%Given that the angle-energy loss correlation does not vanish even in the central region, it should be by far more prominent in the hard scattering region, where the number of scatterings is small, as we know from our experience with Moli\`{e}re and Landau distributions. 
%To be more precise, in the case of Moli\`{e}re distribution, observing an even with large $\Theta$ one can assert only that there was involved only one hard scattering, whereas the number of soft scatterings before and after the hard one can be arbitrary, and generally is large. 
%But observing a joint distribution in $\Theta$ and energy losses $\lambda$, one can say more definitely how many soft scatterings were there. 
%Likewise, in the case of Landau distribution, observing an even with large $\lambda$, one can assert that there occurred a single hard scattering on an atomic electron. The number of soft scatterings on electrons, as well as any scatterings on atomic nuclei, remains uncertain, but it can be determined based on the two-variable distribution. Therefore, even in the hard scattering region, effects of multiple scattering in the 2d distribution are generally not small, and the correlation description must take them into account. 

%Since the correlation intrinsically originates from that in a single-scattering process, it is expectable that in the hard region it must be the least degraded, i.e., be the largest. 

%The impact of .... on ... is that the limits $\Theta\to\infty$ and $\lambda\to\infty$ are non-commutable. Therefore, to be rigorous, we have to consider the two corresponding cases. 

\subsection{Fixed energy loss, large $\Theta$}\label{subsec:fixedlambda-largeTheta}

A simpler case to start with is $\Theta\to\infty$ at $\lambda$ held fixed. 
The corresponding limiting form for $F$ can be read off from Eqs. (\ref{quasi-factorization}) and (\ref{varphi_M-Rutherford}):
\begin{subequations}\label{Fvarphi_LTheta4}
\begin{equation}\label{Fvarphi_LTheta4a}
F(Z,y_0,\Theta,\lambda)
\underset{\Theta\to\infty}\simeq\frac{1}{\pi\Theta^4}\varphi_L(\lambda),
\end{equation}
or
\begin{equation}\label{Fvarphi_LTheta4b}
f\underset{\theta\gg\chi_c}\simeq\frac{2mZ}{\pi p^2\theta^4}\varphi_L\left[\lambda(Z,l,\epsilon)\right].
\end{equation}
\end{subequations}
One corollary is that with the increase of $\Theta$, 
the maximum of $F$ with respect to $\lambda$ approaches $\lambda_m$ for the pure Landau distribution  [$\lambda_m(\Theta)\underset{\Theta\to\infty}\to\lambda_m$, see Fig. \ref{fig:MPEL-Theta}]. 
But insofar as Eq. (\ref{quasi-factorization}) was derived under condition (\ref{lambdallZlny_0}), 
i.e., at limited $\lambda$, there remains an open question whether asymptotics  (\ref{Fvarphi_LTheta4}) is sustained at large $\lambda$.

In fact, it proves to hold for any fixed $\lambda$.
The proof begins with noting that in integral (\ref{f=intdy-intds}), limit $\Theta\to\infty$ at fixed $\lambda$ effectively corresponds to $y\sim\Theta^{-2}\to0$ at fixed $u$. 
The exponential $e^{\Omega_{el}+\Omega_{in}}$ may then be linearized in $\Omega_{el}=-\frac{y}{4}\ln\frac{y_0}{y}$ and $\Omega_{in}(y/4Z,u)-\Omega_{in}(0,u)\underset{y/4Zu\ll1}\simeq\frac{y}{4Z}(2+\ln u)$ [cf. Eq. (\ref{Sigmain-NLO-y})], but not in $\Omega_{in}(0,u)=u\ln u$:
\begin{eqnarray}\label{F-NLO-y}
F(Z,y_0,\Theta,\lambda)
=\frac{1}{2\pi i}\int_{-i\infty}^{i\infty} du e^{\lambda u +u\ln u }\qquad\qquad\qquad\quad
\nonumber\\
\times\frac{1}{4\pi}\int_0^{y_{\max}} dyJ_0(\sqrt{y}\Theta)
\left[-\frac{y}{4}\ln\frac{y_0}{y} +\frac{y}{4Z}(2+\ln u)\right].\quad
\end{eqnarray}
Employing here identity 
\begin{equation}\label{int-Bessel=0}
\int_0^{\infty}dy y J_0(\sqrt{y}\Theta)=0 \qquad (\Theta>0)
\end{equation}
along with relation (\ref{varphi_M-Rutherford}), we are led to asymptotic law (\ref{Fvarphi_LTheta4a}) for any fixed $\lambda$. 
Its actual condition of validity is $\Theta^{2}\gg 1$, and simultaneously $\Theta^{-2}\sim y\ll Zu\sim Z/\lambda$, i.e.,
\begin{equation}\label{Rutherf-cond}
\Theta^{2}\gg \lambda/Z,1. 
\end{equation}

Evidently, the present factorization of the distribution function into Rutherford asymptotics $\theta^{-4}$ times Landau distribution depending on $\lambda$ corresponds to the dominance of events involving one hard scattering on an atomic nucleus (hard scattering on an atomic electron is excluded by requiring the energy transfer to be limited), preceded and followed by statistically independent soft scatterings, whose energy straggling is described by the Landau distribution.

Less trivial, however, is that factorization (\ref{Fvarphi_LTheta4}) is inequivalent to an absence of correlation as expressed by Eq. (\ref{would-be-factoriz}). 
The latter would imply 
\begin{equation}\label{Rutherf-without-correl}
F\approx \frac{Z}{Z+1}\varphi_L(\lambda)\frac{\chi_c^4}{\pi\theta^4}
=\frac{Z+1}{Z}\frac{1}{\pi\Theta^4}\varphi_L(\lambda),
\end{equation}
with an extra factor $\frac{Z+1}{Z}$ including hard scattering on atomic electrons, already mentioned in Sec. \ref{subsec:g-1}. 
Hence, the correlation in this region does not vanish, 
solely because hard scattering on electrons is excluded by demanding $\lambda$ to be limited.

\subsection{Large energy loss}\label{subsec:large-u}

In the limit $\lambda\to\infty$, typical $u$ in the Fourier integral (\ref{f=intdy-intds}) tend to zero. 
That does not allow expanding the exponential $e^{\Omega_{in}}$ in powers of $u$ yet, 
because of its essential nonanalyticity at $u=0$. 
But since the difference $\Omega_{in}(y/4Z,u)-\Omega_{in}(y/4Z,0)$ in this limit is ``almost uniformly" small (see Fig. \ref{fig:Sigmain}),
%\footnote{Strictly speaking, derivative
%\begin{equation}\label{Sigmain's0}
%\end{equation}
%diverges at small and at large $Y$, but only logarithmically. 
%This can be easily compensated by a sufficient smallness of $u$.} 
the exponential may be expanded in the latter difference. 
The nontrivial $u$-integral then greatly simplifies by double partial integration over $u$ with the aid of identity (\ref{d2s=exp}):
\begin{eqnarray}\label{}
F(Z,y_0,\Theta,\lambda)
\underset{\lambda\to\infty}\simeq\frac{1}{4\pi \lambda^2}\qquad\qquad\qquad\qquad\qquad\qquad\nonumber\\
\times\int_0^{y_{\max}} dyJ_0(\sqrt{y}\Theta) e^{\Omega_{el}(y_0,y)+\Omega_{in}\left({y}/{4Z},0\right)}\quad \nonumber\\
\times\frac{1}{2\pi i}\int_{-i\infty}^{i\infty} du e^{\lambda u} \frac{\partial^2}{\partial u^2}\left[\Omega_{in}\left(\frac{y}{4Z},u\right)-\Omega_{in}\left(\frac{y}{4Z},0\right)
\right]\nonumber\\
=\frac{1}{4\pi \lambda^2}\int_0^{y_{\max}} dyJ_0(\sqrt{y}\Theta) e^{\Omega_{el}(y_0,y)+\Omega_{in}\left({y}/{4Z},0\right)}\qquad\nonumber\\
\times\frac{1}{2\pi i}\int_{-i\infty}^{i\infty} \frac{du}{u} e^{\lambda u-{y}/{4Zu}}.\qquad 
\end{eqnarray}
The integral in the last line is recognized to be a zero-order Bessel function \cite{Abr-Steg}:
\[
\frac{1}{2\pi i}\int_{-i\infty}^{i\infty} \frac{du}{u} e^{\lambda u-\frac{y}{4Zu}}=J_0\left(\sqrt{y\lambda/Z}\right),
\]
%where
%\begin{equation}\label{Psi=sqrt}
%\Psi_{\lambda}(\lambda)=\sqrt{}.
%\end{equation}
Combining logarithms in the exponent $\Omega_{el}(y_0,y)+\Omega_{in}\left({y}/{4Z},0\right)$ as in Eq. (\ref{collect-lny+const}), and similarly changing the integration variable from $y$ to $\eta$, we are led to the asymptotic expression
\begin{equation}\label{Fproptog}
F(Z,y_0,\Psi,\lambda)
\underset{\lambda\to\infty}\simeq\frac{Z}{(Z+1)\lambda^2}\beta_M\left(\frac{4\chi_c^2}{\chi'^2_{at}},\Psi,\Psi_{\lambda}\right),
\end{equation}
with
\begin{equation}\label{Psilambda-def}
\Psi_{\lambda}=\sqrt{\frac{\lambda}{Z+1}}
\end{equation}
and
\begin{subequations}\label{betaM}
\begin{eqnarray}
\beta_M(\eta_0,\Psi,\Psi_{\lambda}) \qquad\qquad\qquad\qquad\qquad\qquad\qquad\qquad\quad\nonumber\\
=\frac{1}{4\pi}\int_0^{\eta_{\max}}  d\eta J_0\left(\sqrt{\eta}\Psi\right) J_0\left(\sqrt{\eta}\Psi_{\lambda}\right) e^{-\frac{\eta}{4}\ln\frac{\eta_0}{\eta}}%\nonumber\\
\qquad\quad\,\,\,\,\label{betaM-int-J0J0}\\
=\frac{1}{2\pi}\int d\phi_{\widehat{\bm{\Psi}\bm{\Psi}_{\lambda}}}\varphi_M\left(\eta_0,|\bm{\Psi}-\bm{\Psi}_{\lambda}|\right)\qquad\qquad\qquad\qquad\label{azim-aver-varphiM}\\
=\frac{2}{\pi}\int_{|\Psi-\Psi_{\lambda}|}^{\Psi+\Psi_{\lambda}}\frac{d\Psi'\Psi' \varphi_M(\eta_0,\Psi')}{\sqrt{\left[(\Psi+\Psi_{\lambda})^2-\Psi'^2\right]\left[\Psi'^2-(\Psi-\Psi_{\lambda})^2\right]}}.\nonumber\\
\label{betaM-int-dzzvarphiM-sqrt}
\end{eqnarray}
\end{subequations}
%plays the role of a nuclear rescattering probability distribution. 
Function (\ref{betaM}), symmetrically depending on its arguments $\Psi$ and $\Psi_{\lambda}$, is normalized by
\begin{eqnarray}\label{}
2\pi\int_0^{\infty}d\Psi\Psi \beta_M(\eta_0,\Psi,\Psi_{\lambda})\qquad\qquad\quad\nonumber\\
=2\pi\int_0^{\infty}d\Psi_{\lambda}\Psi_{\lambda} \beta_M(\eta_0,\Psi,\Psi_{\lambda})=1.
\end{eqnarray}
It is also encountered in other multiple Coulomb scattering problems \cite{Weber-Bell,Jakas}.

Distribution (\ref{Fproptog}), (\ref{azim-aver-varphiM}) describes the probability of a two-stage process: a single hard scattering on an atomic electron through an angle $\Psi_{\lambda}$, 
whose modulus corresponds to the energy loss $\lambda$, 
and an accompanying multiple elastic and inelastic scattering from $\bm{\Psi}_{\lambda}$ to angle $\bm\Psi$ with a prescribed modulus $\Psi$. 
Factor $\lambda^{-2}$ in (\ref{Fproptog}) is nothing but the high-$\lambda$ asymptotics of $\varphi_L(\lambda)$ [cf. Eq. (\ref{varphiL-Ruth})]. 
Compared to (\ref{quasi-factorization}), however, the dependence on $\lambda$ in the factorized structure (\ref{Fproptog}) at large $\lambda$ enters not to the thickness parameter $y_0$, but to the angular variable $\Psi_{\lambda}$ on a par with $\Psi$. 
The accuracy of this approximation is illustrated in Figs. \ref{fig:PlotZ1}b,c, by dot-dashed curves, showing that it works well for $\lambda\gtrsim 20$.

\begin{figure}
\includegraphics{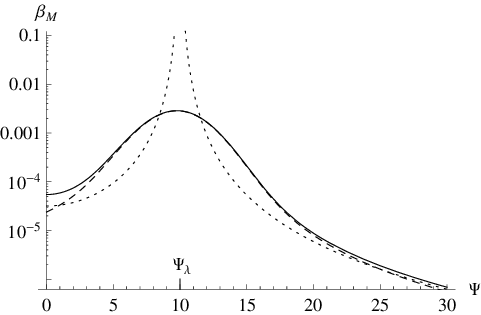}
\caption{\label{fig:phiM-Theta-Psi} Multiple Coulomb rescattering probability distribution (\ref{betaM-int-J0J0}) in the spur region, for $\eta_0=10^4$, $\Psi_{\lambda}=10$ [$\lambda=100(Z+1)$]. 
Dashed curve, large-$\Psi_{\lambda}$ approximation (\ref{g-ridge}), (\ref{projected-varphiM}). 
Dotted, double hard scattering approximation (\ref{g-algebr}).
}
\end{figure}

We are primarily interested in the behavior of $\beta_M$ at large $\Psi_{\lambda}$. 
There, it admits further simplifications. 
In particular, if $|\Psi-\Psi_{\lambda}|$ is substantially smaller than $\Psi_{\lambda}$ (not too far from the midline of the spur), one can neglect in the denominator of (\ref{betaM-int-dzzvarphiM-sqrt}) $\Psi'^2$ compared with $(\Psi+\Psi_{\lambda})^2$, but not compared with $(\Psi-\Psi_{\lambda})^2$. That leads to an approximation
\begin{equation}\label{g-ridge}
\beta_M(\eta_0,\Psi,\Psi_{\lambda})\simeq\frac{1}{\pi(\Psi+\Psi_{\lambda})}\varphi_{Mx}\left(\eta_0,\Psi-\Psi_{\lambda}\right),
\end{equation}
where
\begin{equation}\label{projected-varphiM}
\varphi_{Mx}\left(\eta_0,\Psi_x\right)=\frac{1}{\pi}\int_0^{\sqrt{\eta_{\max}}}dx\cos \left(\Psi_x x\right)  e^{-\frac{x^2}{4}\ln\frac{\eta_0}{x^2}}
\end{equation}
is the projection of the Moli\`{e}re distribution onto one of the two Cartesian components of transverse vector $\bm\Psi$ \cite{Bond-Moliere} [the averaging over a circle $|\bm\Psi|=\Psi$ in Eq. (\ref{azim-aver-varphiM}) is replaced by integration over the line tangential to it in a point $\bm\Psi=\bm{\Psi}_{\lambda}(\bm\Psi\cdot\bm\Psi_{\lambda})/\Psi_{\lambda}^2$ nearest to $\bm{\Psi}_{\lambda}$].
This approximation is confronted with the more accurate result (\ref{Fproptog}), (\ref{betaM-int-J0J0}) in Fig. \ref{fig:phiM-Theta-Psi}. 
Its accuracy is good near the top of the distribution, 
but degrades away from it, where $(\Psi-\Psi_{\lambda})^2$ is no longer small compared with $(\Psi+\Psi_{\lambda})^2$.

If $\Psi_{\lambda}$ and $\Psi$ are about equally large but not close, 
the argument of $\varphi_M$ in the integrand of (\ref{azim-aver-varphiM}) or (\ref{betaM-int-dzzvarphiM-sqrt}) is everywhere large, as well. 
Then, pure Rutherford asymptotics (\ref{varphi_M-Rutherford}) applies on the entire integration interval, 
and the integral is simple to evaluate: 
\begin{equation}\label{g-algebr}
\beta_M(y_0,\Psi,\Psi_{\lambda})\underset{\Psi,\Psi_{\lambda},|\Psi-\Psi_{\lambda}|\gg1}\simeq
\frac{\Psi^2+\Psi_{\lambda}^2}{\pi \left|\Psi^2-\Psi_{\lambda}^2\right|^3}.
\end{equation}
One infers from here that if $\Psi$ is held fixed and $\Psi_{\lambda}$ is sent to infinity, 
function $\beta_M$ decreases as $\beta_M\underset{\Psi_{\lambda}\to\infty}\sim\Psi_{\lambda}^{-4}\sim \lambda^{-2}$, 
which, according to Eq. (\ref{Fproptog}), corresponds to the large-$\lambda$ asymptotic behavior for $F$: 
\begin{equation}\label{F-large-u}
F\underset{\lambda\to\infty}\sim \lambda^{-4}.
\end{equation}
Besides that, Eq. (\ref{g-algebr}) predicts a Rutherford-like asymptotics at $\Psi\to\infty$. 
But insertion of (\ref{g-algebr}) to Eq. (\ref{Fproptog}) specifically yields
\begin{equation}\label{}
F\underset{\Psi\to\infty}
\simeq
\frac{Z}{(Z+1)\pi \lambda^2 \Psi^4}
=\frac{Z+1}{Z \pi \lambda^2 \Theta^4},
\end{equation}
which due to the $\frac{Z+1}{Z}$ factor matches with (\ref{Rutherf-without-correl}), corresponding to the absence of correlation, and somewhat differing from the correct asymptotics (\ref{Fvarphi_LTheta4}). 
Hence, at sufficiently large $\Psi$ approximation (\ref{Fproptog}) breaks down. That is also confirmed by Fig. \ref{fig:PlotZ1}b.

%If, on the contrary, $\Psi=\Psi_{\lambda}$, (\ref{azim-aver-varphiM}) reduces to
%\[
%\frac{2}{\pi}\int_0^{2\Psi}\frac{dz}{\sqrt{4\Psi^2-z^2}}\varphi_M(y_0,z)
%\underset{\Psi\to\infty}\simeq
%\frac{1}{\pi\Psi}\int_0^{\infty}dz\varphi_M(y_0,z),
%\]
%or
%\[
%\frac{1}{2\pi}\int_0^{\sim B_0}dBB J_0^2(B\Psi)e^{-\frac{B^2}{2}\ln\frac{B_0}{B}}
%\underset{\Psi\to\infty}\simeq
%\frac{1}{2\pi^2\Psi}\int_0^{\sim B_0}dB e^{-\frac{B^2}{2}\ln\frac{B_0}{B}}.
%\]

\section{Conditional mean values}\label{sec:mean-values}

In the previous section, studying the correlation in the 2-variable distribution function, we had seen that the correlation between $\Theta$ and $\lambda$ strengthens at large values of those variables. 
But the quadratic falloff of the probability density with $\lambda$ can hamper experimental observation of  correlation effects in this region. 
In view of that, advantageous may be experimental setups directly measuring net characteristics of the distribution, such as conditional mean values and widths. 
The corresponding momentum or energy weighting factors enhance the relative contribution of the hard component. 
In what follows, we will investigate manifestations of the correlation directly for such net quantities.

\subsection{Energy loss dependence of angular dispersion}\label{subsec:6A}

\begin{figure}
\includegraphics{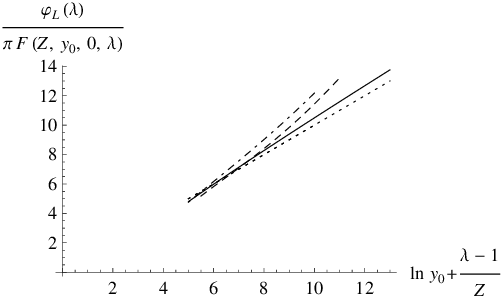}
\caption{\label{fig:meansqTheta-lin-growth} Inverse of the angular distribution function at zero deflection angle, serving as a counterpart of the mean square deflection angle. 
Dot-dashed curve, $y_0=10^3$, $Z=1$. 
Dashed curve, $y_0=10^3$, $Z=3$. 
Solid, $y_0=10^6$, $Z=1$. This curve also merges with the generic $y_0K$-scaling approximation. 
Dotted line, leading logarithmic approximation (\ref{lny0+u/Z}).
}
\end{figure}

The shape of the angular distribution at fixed $\lambda$, at least when $\lambda$ is moderate, is symmetric and bell-like, so, the measure of its width could be the mean square deflection angle. 
But Rutherford asymptotics of the distribution at large $\Theta$ leads to a logarithmic divergence of $\left\langle\Theta^2\right\rangle$ for any $\lambda$. 
%Nonetheless, for moderate $\lambda$, the $\Theta$-distribution in the central region is close to a Gaussian, wherefore there must exist some proper characteristic of its width, anyway.
%The results of experimental measurements are often fitted by a Gaussian.
As a finite counterpart of $\left\langle\Theta^2\right\rangle$, 
one can merely take the inverse of the distribution function in the origin. %This is a finite quantity, simply related with the distribution function. 
According to Eq. (\ref{Moliere-LLA}),
%\begin{subequations}\label{LLA-ang-disp}
\begin{equation}\label{LLA-ang-disp-y0K}
\frac{\varphi_L(\lambda)}{\pi F(Z,y_0,0,\lambda)}\approx \frac{1}{\pi\varphi_M(y_0K,0)},
\end{equation}
i.e., all the $\lambda$, $Z$ and $y_0$ dependencies reduce to a dependence on a single variable $y_0K(Z,\lambda)$. 
This scaling holds only if $\ln y_0$ is large enough, 
so that condition (\ref{lambdallZlny_0}) can be met in spite that $\lambda/Z\gtrsim1$. 
From Fig.~\ref{fig:meansqTheta-lin-growth} it is evident that at $y_0=10^3$ ($\ln y_0\approx 7$) and low $Z$ the scaling is somewhat violated. 
When the scaling does hold, one can furthermore employ in (\ref{LLA-ang-disp-y0K}) the leading logarithmic approximation (\ref{Moliere-LLA}) to get
\begin{equation}\label{lny0+u/Z}
\frac{\varphi_L(\lambda)}{\pi F(Z,y_0,0,\lambda)}\approx \ln y_0K=\ln y_0+\frac{\lambda-1}{Z}.
\end{equation}
%\end{subequations}
Approximation (\ref{lny0+u/Z}), slowly improving with the increase of $y_0$, rises linearly with $\lambda$, provided the $\lambda$-dependent contribution is relatively small, satisfying condition (\ref{lambdallZlny_0}). 
This linear law may be thought of as a sum of soft ($\ln y_0$) and hard ($\frac{\lambda-1}{Z}$) contributions. 
Curiously, as Fig. \ref{fig:meansqTheta-lin-growth} indicates, 
the linearity is sustained even when the scaling is violated (dashed and dot-dashed curves), 
but the slope of the $\lambda$ dependence then becomes dependent on $y_0$. 
It is also clear that at sufficiently large $\lambda$ the distribution shape becomes non-Gaussian, and (\ref{LLA-ang-disp-y0K}) can no longer serve as a measure of the mean squared deflection angle.

%Instead, the latter becomes strongly influenced by location $\Theta\approx \Psi=\sqrt{\lambda/Z}$ of the slice of the spur. 
%Squaring this, we see that the effective $\langle\Theta^2\rangle$ continues growing with $\lambda/Z$ linearly.

\subsection{Angular dependence of the mean energy loss}\label{subsec:barlambda(Theta)}

In what concerns the energy straggling distribution, which is highly asymmetric, its simplest characteristic is the mean energy loss. 
Historically, magnetospectrometric measurements of this quantity at fixed angles for nonrelativistic incident ions were the first to give experimental evidence for an angle-energy loss correlation. 
Magnetic spectrometry is applicable for relativistic electrons, too \cite{Weigold-McCarthy}. 
%Its accuracy is rather high, and may permit measurement at least of  the mean energy loss. 
Alternatively, one may consider using a sufficiently large solid-state detector, registering events up to very high $\lambda$, summation over which can give the mean ionization energy loss at a fixed deflection angle. 
%For relativistic electrons, the spectrometric method must be different, due to the necessity to eliminate radiative losses, but should be within experimental reach, as well (see Introduction). 

Since energy loss is linearly related to the reduced variable $\lambda$ [see Eq. (\ref{u-energyloss})], 
their mean values are related linearly, as well:
\begin{equation}\label{meanetheta-meanuTheta}
\overline{\epsilon}(\theta)
=\frac{\int_0^{\infty} d\epsilon\epsilon  f(\theta,\epsilon)}{\int_0^{\infty} d\epsilon  f(\theta,\epsilon)}%\nonumber\\
=\frac{p^2Z\bar\chi_c^2}{2m}\left[\lambda_0+\bar\lambda(\Theta)\right],%\qquad 
\end{equation}
where $\lambda_0$ is defined by Eq. (\ref{lambda0-def}), and
\begin{equation}\label{meanuTheta-def}
\bar{\lambda}(\Theta)=\frac{\int_{-\infty}^{\infty} d\lambda\lambda F(Z,y_0,\Theta,\lambda)}{\int_{-\infty}^{\infty} d\lambda F(Z,y_0,\Theta,\lambda)}.
\end{equation}
As in previous sections, 
the lower limit of both $\lambda$ integrals was extended from $-\lambda_0$ to $-\infty$ by virtue of the rapid convergence.
At the upper limit, the convergence of the integrals is ensured by asymptotics (\ref{F-large-u}). 
This is in contrast with $\int^{\infty}d\lambda\lambda\varphi_L(\lambda)$, 
which diverges logarithmically because $\varphi_L$ has a slower asymptotics (\ref{varphiL-Ruth}) 
(at extremely large $\lambda$, Landau distribution itself is invalid, 
and the physical, Bethe-Bloch mean energy loss, of course, is finite). 
Although $\lambda$-dependence of $F$ is similar to that of $\varphi_L(\lambda)$ at moderate $\lambda$, 
beyond location of the spur that similarity is violated, 
and the decrease steepens.
%Granted that function $F$ decreases at large $\lambda$ faster than $\varphi_L$ [see Eqs. (\ref{F-large-u}), (\ref{varphiL-Ruth})], the integral in the numerator of (\ref{meanuTheta-def}) converges. 

%Whereas for Landau distribution, possessing asymptotics (\ref{varphiL-Ruth}), integral $\int^{\infty}d\lambda\lambda\varphi_L(\lambda)$ logarithmically diverges at the upper limit 
%(implying that to evaluate the physical mean energy loss, one needs to take into account its limitation by the maximal kinematically allowed energy transfer $\epsilon_{\max}(E)=E-mc^2$ including previously neglected longitudinal motion and relativistic effects for secondary electrons), a similar integral $\int_{-\infty}^{\infty} d\lambda\lambda  F(Z,y_0,\Theta,\lambda)$ at any fixed $\Theta$ converges, because according to Eq. (\ref{F-large-u}), $F$ decreases more rapidly than $\varphi_L$. 
%But asymptotics (\ref{F-large-u}) only sets in beyond the ``spur'', i.e., at values greater than
%\begin{equation}\label{lambdamax-Theta}
%\lambda_{\max}(\Theta)\sim Z\Theta^2= (Z+1)\Psi^2
%\end{equation}
%[see Eqs. (\ref{lambda=ZTheta2}), (\ref{Psi-def})]. 
%So, in effect, this spur furnishes an upper cutoff $\lambda_{\max}(\Theta)$, which is smaller and more relevant than that imposed by $\epsilon_{\max}(E)$. 

Our task now is to evaluate ratio (\ref{meanuTheta-def}) 
by substituting there the integral representation (\ref{f=intdy-intds}) for $F$. 
With the use of identity (\ref{intinftydlambda}) and its derivative
\[
\int_{-\infty}^{\infty} d\lambda\lambda\frac{1}{2\pi i}\int_{-i\infty}^{i\infty}du e^{\lambda u+\Omega_{in}(Y,u)}=-\frac{\partial}{\partial u}e^{\Omega_{in}(Y,u)}\bigg|_{u=0},
\]
we recast (\ref{meanuTheta-def}) as
\begin{eqnarray}\label{barlambda-through-dSigmain}
\bar{\lambda}(\Theta)=-\left[\int_0^{y_{\max}} dy J_0(\sqrt{y}\Theta) e^{-\frac{y}{4}\ln\frac{y_0}{y}}\right]^{-1}\qquad\qquad\nonumber\\
\times \int_0^{y_{\max}} dy J_0(\sqrt{y}\Theta) e^{\Omega_{el}(y_0,y)+\Omega_{in}(y/4Z,0)}\nonumber\\
\times \frac{\partial}{\partial u}\Omega_{in}(y/4Z,u)\bigg|_{u=0}.\qquad\quad
\end{eqnarray}
Substituting 
$\frac{\partial}{\partial u}\Omega_{in}(Y,u)\big|_{u=0}=\ln Y+1+\gamma_{\text{E}}$, 
and passing to integration variable $\eta$ [Eq. (\ref{eta-def})],
we transform (\ref{barlambda-through-dSigmain}) to
\begin{subequations}\label{barlambdaTheta-abc}
\begin{eqnarray}\label{mean-u(Theta)-def}
\bar{\lambda}(Z,\eta_0,\Psi)
%&=&\frac{\int_{-\infty}^{\infty} d\lambda\lambda F(Z,y_0,\Theta,\lambda)}{\int_{-\infty}^{\infty} du F(Z,y_0,\Theta,\lambda)}\nonumber\\
%&=&-\frac{\int_0^{\sim y_0} dy J_0(\sqrt{y}\Theta) e^{-\frac{y}{4}\ln\frac{y_0}{y}} \frac{\partial}{\partial s}\left[s\ln u-\Delta(y,u)\right]\Big|_{s=0}}{\int_0^{\sim y_0} dy J_0(\sqrt{y}\Theta) e^{-\frac{y}{4}\ln\frac{y_0}{y}}}\nonumber\\
&=&\frac{\int_0^{\eta_{\max}} d\eta J_0(\sqrt{\eta}\Psi) e^{-\frac{\eta}{4}\ln\frac{\eta_0}{\eta}} \ln\frac{1}{\eta}}{\int_0^{\eta_{\max}} d\eta J_0(\sqrt{\eta}\Psi) e^{-\frac{\eta}{4}\ln\frac{\eta_0}{\eta}}}\nonumber\\
&\,&+\ln 4(Z+1)-1-\gamma_{\text{E}}, 
%&\equiv&-\frac{\int_0^{\sim \eta_0} d\eta J_0(\sqrt{\eta}\Psi) e^{-\frac{\eta}{4}\ln\frac{\eta_0}{\eta}} }{\pi \varphi_M(\eta_0,\Psi)}\nonumber\\
%&\,&\qquad\qquad
%+\ln \frac{4(Z+1)}{\eta_0}-\gamma_{\text{E}},
%\end{eqnarray}
%Integrating by parts, we can also rewrite it as
%\begin{eqnarray}\label{}
%\bar{\lambda}(Z,y_0,\Theta)
\end{eqnarray}
where $\Psi$ and $\eta_0$ are given by Eqs. (\ref{Psi-def}), (\ref{eta0-def}).
The integral in the denominator is recognized to equal $4\pi\varphi_M(\eta_0,\Psi)$. 
Representation (\ref{meanetheta-meanuTheta}), (\ref{mean-u(Theta)-def}) can as well be derived directly from the generic solution (\ref{double-Fourier}), (\ref{fblambda=expJ0e}) of the transport equation (see Appendix \ref{app:Jakas}).

%\newpage

\begin{figure}
\includegraphics{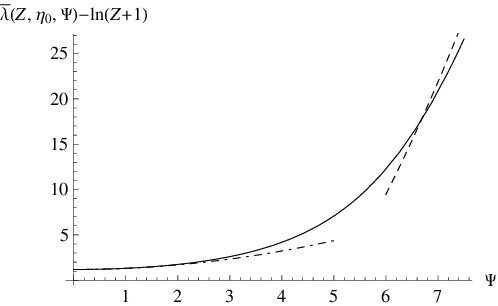}
\caption{\label{fig:mean-mean-u(Theta)} Solid curve, dependence of the mean energy loss defined by Eq. (\ref{mean-u(Theta)-def}) on the scattering angle, at $\eta_0=10^4$. 
Dot-dashed curve, approximation (\ref{bar-u-low-theta}). 
Dashed curve, approximation (\ref{baru=theta2-lntheta+const}). }
\end{figure}

In fact, the numerator of (\ref{mean-u(Theta)-def}), too, can be expressed through $\varphi_M$. 
To this end, it is expedient first to integrate in the numerator by parts and isolate there all the terms proportional to $\varphi_M$, which cancels with $\varphi_M$ in the denominator:
\begin{eqnarray}\label{baruTheta-byparts-new}
\bar{\lambda}(Z,\eta_0,\Psi)&=&-\frac{\int_0^{\eta_{\max}} d\eta J_0(\sqrt{\eta}\Psi) \frac{\partial}{\partial\eta} e^{-\frac{\eta}{4}\ln\frac{\eta_0}{\eta}}}{\pi \varphi_M(\eta_0,\Psi)}\qquad \nonumber\\
&\,&
-\ln \frac{\eta_0}{4(Z+1)}-\gamma_{\text{E}},
\end{eqnarray}
where
\[
\ln \frac{\eta_0}{4(Z+1)}+\gamma_{\text{E}}=\frac{Z+1}{Z} \left(\ln\frac{Z\bar\chi_c^2}{\chi'^{2}_{at} }+\gamma_{\text{E}}\right).
\]
That eliminates in the integrand the factor $\ln\frac{1}{\eta}$, which is absent in integral (\ref{Moliere-def}) defining $\varphi_M$. 
Now, once again integrating in (\ref{baruTheta-byparts-new}) by parts, employing identity
\begin{equation*}\label{}
\frac{\partial}{\partial\eta}J_0(\sqrt{\eta}\Psi)
=-\frac12 \int_0^{\Psi}d\Psi'\Psi' J_0(\sqrt{\eta}\Psi')
\end{equation*}
[an integral form of the Bessel equation $\frac{d}{d\xi}\xi\frac{d}{d\xi}J_0(\xi)=-\xi J_0(\xi)$, $J'_0(0)=0$], and interchanging the order of $\eta$- and $\Psi$-integrations,
%using 
%\[
%-\frac{dJ_0}{dz}=\frac{1}{z}\int dzz J_0(z),
%\]
%\begin{eqnarray}\label{baruTheta-byparts}
%\bar{\lambda}(Z,\eta_0,\Psi)&=&\frac{1}{\pi \varphi_M(\eta_0,\Psi)}
%-\ln \frac{\eta_0}{4(Z+1)}-\gamma_{\text{E}}\qquad \nonumber\\
%&\,&-\frac{\theta}{\chi_c}\frac{\int_0^{\sim \eta_0} \frac{d\eta}{\sqrt{\eta}} J_1(\sqrt{\eta}\Psi) e^{-\frac{\eta}{4}\ln\frac{\eta_0}{\eta}}}{2\pi \varphi_M(\eta_0,\Psi)},
%\end{eqnarray}
%where
%\[
%\ln \frac{\eta_0}{4(Z+1)}+\gamma_{\text{E}}=\frac{Z+1}{Z} \left(\ln\frac{Z\bar\chi_c^2}{\chi'^{2}_{at} }+\gamma_{\text{E}}\right).
%\]
%In the numerator of the last term of (\ref{baruTheta-byparts}), we can now substitute the inverse of Eq. (\ref{Moliere-def}),
%\[
% e^{-\frac{\eta}{4}\ln\frac{\eta_0}{\eta}}=2\pi \int_0^{\infty}d\Psi\Psi J_0(\sqrt{\eta}\Psi)\varphi_M(\eta_0,\eta),
%\]
%and use identity 
%\[
%\int_0^{\infty}\frac{d\eta}{\sqrt{\eta}}J_1(\sqrt{\eta}\Psi)J_0(\sqrt{\eta}\Psi)=....
%\]
%to 
one expresses the angular dependence of the mean energy loss through the Moli\`{e}re function $\varphi_M$ alone:
\begin{equation}\label{intvarphiM/varphiM}
\bar{\lambda}(Z,\eta_0,\Psi) =\frac{2\int_{\Psi}^{\infty}d\Psi'\Psi' \varphi_M(\eta_0,\Psi')}{ \varphi_M(\eta_0,\Psi)}
-\ln \frac{\eta_0}{4(Z+1)}-\gamma_{\text{E}}.
\end{equation}
\end{subequations}
In contrast to representation (\ref{Jakas}), it is explicitly independent of the underlying single-scattering differential cross section, except the dependence on $\chi'_{at}$ entering $\eta_0$. 
It appears also that for a given $\eta_0$, the $\Psi$-dependent part of $\bar\lambda$ is independent of $Z$. 
That property may obviate the need to use in experiments only lowest-$Z$ target materials.

To explicate the $\Psi$ dependence of $\bar{\lambda}(Z,\eta_0,\Psi)$, it is appealing to employ 
approximations for $\varphi_M$ obtained in Sec. \ref{subsec:Moliere}. 
However, the leading logarithmic approximation (\ref{Moliere-LLA}) is not suitable for that purpose, 
insofar as the ratio represented by the first term in Eq. (\ref{intvarphiM/varphiM}) then equals $\ln\eta_0$, being $\Psi$-independent and exactly canceling even the similar $\eta_0$-dependence in the second term. 
Therewith
\[
\bar{\lambda}(Z,\eta_0,\Psi) 
\approx \ln 4(Z+1) -\gamma_{\text{E}}.
\]
depends only on $Z$ due to the existence of the effective $Z$-dependent cutoff at  the spur. 
One thus needs more accurate approximations for $\varphi_M$. 
It is clear also that at small and at large $\Psi$ the approximations should differ. 
Let us begin with the small $\Psi$ region.
%In reality, though, as is illustrated in Fig. \ref{fig:mean-mean-u(Theta)},  function $\bar{\lambda}(Z,\eta_0,\Psi)$ monotonously increases with $\Psi$. 
%We will show below that its $\Psi$ dependence is quadratic both at $\Psi\to0$ and $\Psi\to\infty$, but with different coefficients. 

\begin{figure}
\includegraphics{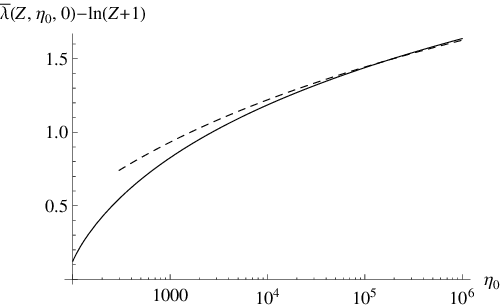}
\caption{\label{fig:u(0)vsy0} Target thickness dependence of the mean energy loss in the forward direction. 
Solid curve, exact result [Eq. (\ref{bar-u(0)})]. 
Dashed, approximation (\ref{bar-lambda=loglogeta0}).
 }
\end{figure}

\subsubsection{Small $\Psi$}

The minimal value of the monotonously rising function $\bar{\lambda}(Z,\eta_0,\Psi)$, achieved at $\Psi=0$, is
\begin{equation}\label{bar-u(0)}
\bar{\lambda}(Z,\eta_0,0)
=\frac{1}{\pi \varphi_M(\eta_0,0)}
-\ln \frac{\eta_0}{4(Z+1)}-\gamma_{\text{E}}.
\end{equation}
This value is positive  (see Fig. \ref{fig:u(0)vsy0}), 
and thus exceeds the most probable energy loss [cf. Eq. (\ref{lambdam})]. 
That is natural in view of the high skewness of the $\lambda$-distribution. 
Employing in (\ref{bar-u(0)}) approximation (\ref{1piMolierePsi0-NLLA}), we get
\begin{equation}\label{bar-lambda=loglogeta0}
\bar{\lambda}(Z,\eta_0,0)
\underset{\ln\eta_0\gg1}\simeq\ln\left[(Z+1)\ln\eta_0\right]-1,
\end{equation}
implying a very slow thickness dependence.
%It also depends on $\eta_0$, but slower than logarithmically, due to a cancellation of $\ln \eta_0$ in the right-hand side of (\ref{bar-u(0)}).
%In the LLA, that would equal $\ln 4(Z+1)-\gamma_{\text{E}}$, being virtually independent of the target thickness represented by $\eta_0$ [with $\bar\epsilon(\theta=0)$ yet depending on $l$ via Eq. (\ref{lambda-def})], but Fig. \ref{fig:u(0)vsy0} shows that dependence of the straggling on $\eta_0$ remains, albeit being slower than logarithmic. 

\begin{figure}
\includegraphics{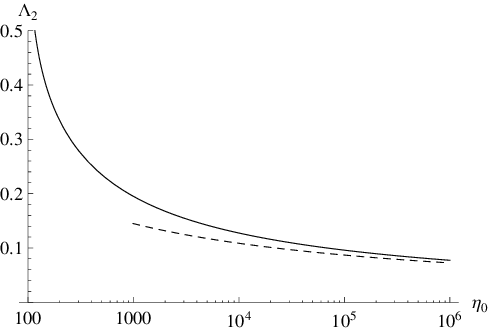}
\caption{\label{fig:u''(0)vsy0} Target thickness dependence of the coefficient at the quadratic term in the $\bar{\lambda}$ dependence on $\Psi$ at small $\Psi$. 
Solid curve, exact result [Eqs. (\ref{bar-u-low-theta}), (\ref{mu2-expr})]. 
Dashed, approximation (\ref{Lambda2=1logeta0}).
 }
\end{figure}

The $\mathcal{O}(\Psi^2)$ correction to (\ref{bar-u(0)}) has the form
\begin{equation}\label{bar-u-low-theta}
\bar{\lambda}(Z,\eta_0,\Psi)
\underset{\Psi \to0}\simeq
\bar{\lambda}(Z,\eta_0,0)+ \Lambda_2(\eta_0)\Psi^2,
\end{equation}
with the coefficient at the quadratic term
\begin{eqnarray}\label{mu2-expr}
\Lambda_2(\eta_0)=\frac{\partial \bar{\lambda}}{\partial \Psi^2}\bigg|_{\Psi=0}
=\frac{\partial}{\partial \ln\eta_0}\bar{\lambda}(Z,\eta_0,0)\nonumber\\
=\frac{\partial}{\partial \ln\eta_0}\frac{1}{\pi\varphi_M(\eta_0,0)}-1,
\end{eqnarray}
where in derivation of the second equality we used Eq. (\ref{mean-u(Theta)-def}), with $\partial J_0(\sqrt{\eta}\Psi)/\partial\Psi^2\big|_{\Psi=0}=-\eta/4$, 
while in the third equality, Eq. (\ref{intvarphiM/varphiM}). 
%\[
%\bar{\lambda}''_\Theta(Z,y_0,0)=
%2\left[
%\frac{\partial}{\partial \ln y_0}\frac{1}{\pi\varphi_M(y_0,0)}
%-1\right].
%\]
In the leading logarithmic approximation for $\varphi_M$ 
[see Eq. (\ref{Moliere-LLA})], the difference in the right-hand side of (\ref{mu2-expr}) would equal zero, similarly to the conclusion reached in \cite{Sigmund-Winterbon} for ions. 
But  employing a more accurate Eq. (\ref{1piMolierePsi0-NLLA}), we find
\begin{equation}\label{Lambda2=1logeta0}
\Lambda_2\underset{\ln\eta_0\gg1}\simeq\frac{1}{\ln\eta_0}.
\end{equation}
Given its slow decrease rate, it remains sizable even at very large target thicknesses. 
The accuracy of approximation (\ref{Lambda2=1logeta0}) is illustrated by Fig. \ref{fig:u''(0)vsy0}.

\subsubsection{Large $\Psi$}

An approximate quadratic $\Psi$-dependence of 
$\bar\lambda(Z,\eta_0,\Psi)$ is also found in the hard region, 
but with a larger coefficient. 
Employing the known expansion \cite{Bond-Moliere}
\begin{equation}\label{}
\pi \varphi_M(\eta_0,\Psi)\underset{\Psi\to\infty}\simeq
\frac{1}{\Psi^4}+\frac{4}{\Psi^6}\left(\ln\frac{\eta_0\Psi^2}{4}+2\gamma_{\text{E}}-3\right)
\end{equation}
(including a correction to the Rutherford asymptotics of $\varphi_M$ at high $\Psi$, with ${\eta_0\Psi^2}/{4}=\theta^2/\chi'^2_{at}$),
integrating it as
\begin{eqnarray*}
2\pi \int_{\Psi}^{\infty}d\Psi'\Psi' \varphi_M(\eta_0,\Psi')
\qquad\qquad\qquad\qquad\qquad\nonumber\\
\underset{\Psi\to\infty}\simeq \frac{1}{\Psi^2}+\frac{4}{\Psi^4}\left(\frac12\ln\frac{\eta_0\Psi^2}{4}+\gamma_{\text{E}}-\frac54\right),
\end{eqnarray*}
and inserting to representation (\ref{intvarphiM/varphiM}), 
we get
\begin{eqnarray}\label{baru=theta2-lntheta+const}
\bar{\lambda}(Z,\eta_0,\Psi)\underset{\Psi\to\infty}\simeq\Psi^2
\frac{1+\frac{4}{\Psi^2}\left(\frac12\ln\frac{\eta_0\Psi^2}{4}+\gamma_{\text{E}}-\frac54\right)}{1+\frac{4}{\Psi^2}\left(\ln\frac{\eta_0\Psi^2}{4}+2\gamma_{\text{E}}-3\right)}\quad\nonumber\\
-\ln \frac{\eta_0}{4(Z+1)}-\gamma_{\text{E}}\quad\nonumber\\
\underset{\Psi\to\infty}\simeq 
\Psi^2-4\ln\frac{\Psi}{2} -\ln \frac{\eta_0^3}{4(Z+1)}-5\gamma_{\text{E}}+7.\qquad
\end{eqnarray}
This asymptote is plotted in Fig. \ref{fig:mean-mean-u(Theta)} by the dashed curve.

The coefficient at the quadratic term in (\ref{baru=theta2-lntheta+const}) is about an order of magnitude greater than in (\ref{bar-u-low-theta}). 
Also, it does not depend on $\eta_0$ (target thickness). 
That is natural, since physically it corresponds to a single hard scattering on an atomic electron. 
But compared with relationship (\ref{Psilambda-def}) between energy transfer and deflection angle in a hard scattering event, in (\ref{baru=theta2-lntheta+const}) there is no factor $Z+1$. 
That is explained by noting that in contrast to $\lambda$ in Eq. (\ref{Psilambda-def}), $\bar\lambda$ in Eq. (\ref{baru=theta2-lntheta+const}) is the mean energy loss. 
Scattering on an electron through a large angle $\Psi$ does give a contribution to $\lambda$ equal $(Z+1)\Psi^2$, 
but the probability of scattering through a large angle on an electron, 
and not on a nucleus (which does not change $\lambda$), is $(Z+1)^{-1}$. 
As a result, those $Z$-dependent factors cancel.

More intricate is the correction term in (\ref{baru=theta2-lntheta+const}), logarithmically depending on $\Psi$. Structurally, it corresponds to a double hard scattering \cite{Bond-Moliere} -- on an electron and on an atomic nucleus, but enters with a negative coefficient. 
That can be explained as follows. 
In an event containing one hard inelastic scattering act, there is a contribution to the aggregate deflection angle from elastic scattering, 
which adds incoherently to the deflection angle squared. 
Correspondingly, the scattering angle squared acquired only in the hard collision with an atomic electron is smaller than that observed, 
implying a negative correction to $\Psi^2$ in (\ref{baru=theta2-lntheta+const}).

%At not too large $\Psi$, pretty large $\bar\lambda$ as compared with MPEL. Comparison must be made not with MPEL but with $\bar\lambda(0)$.

\section{Summary}

The unification of Moli\`{e}re and Landau theories presented herein reveals a pronounced correlation between the deflection angle and ionization energy loss for fast electrons or positrons traversing amorphous matter.
The correlation arises at the single-scattering level, 
and is not quickly ruined by multiple Coulomb scattering, 
due to the anomalous character of the latter. 
The target thickness dependence of the correlation effect is mild. 
As for its $Z$-dependence, the angular distribution in the central region (moderate $\Theta$ and $\lambda$) predominantly depends on the ratio of the straggling variable $\lambda$ to $Z$ 
(see Fig. \ref{fig:meansqTheta-lin-growth}). 
Therefore, even for high-$Z$ target materials, in principle, there are manifestations of the correlation at proportionally high $\lambda$, but they are obscured by a falloff of the event rate with the increase of $\lambda$. 
%, but its experimental observability becomes more difficult at large $Z$, because even though the correlation remains there, as well [being $\mathcal{O}(\frac{1}{Z+1})$ -- see Figs. \ref{fig:g-1}b, \ref{fig:meansqTheta-lin-growth}], it is concentrated at energy losses much higher than most probable. 
%That might be reason why it did not reveal itself in numerous experiments thus far. 
At high $Z$, an observable sensitive to the correlation is the mean energy loss as a function of the deflection angle, whose variable part, representing  the correlation effect, is independent of $Z$ at all (see Sec. \ref{subsec:barlambda(Theta)}). 
The range of applicability of the present theory is basically an intersection of ranges of applicability of Moli\`{e}re and Landau theories, being rather broad.

It may be worth recapitulating that the hard incoherent scattering mechanism responsible for the angle-energy loss correlation for fast electrons differs from those at work for slow ions, quoted in the Introduction. 
Whereas interaction of a nonrelativistic ion with an atom is intrinsically semiclassical, when each impact parameter corresponds to a well-defined scattering angle, as well as energy loss, for relativistic electrons the angle and the energy loss are distributed statistically. 
Nonetheless, their mutual correlation is not small, because it is dominated by hard electron-electron scattering, 
in which the deflection angle and energy transfer are interrelated kinematically, 
irrespective of the impact parameter relative to the atomic nucleus. 
Precisely this contribution survives under multiple scattering conditions. 

At the same time, the hard incoherent scattering correlation shares some properties with other correlation mechanisms.
%For physical interpretation of our results and their comparison with results for incident ions obtained earlier, it is worth emphasizing the picture brought out by our calculations.
%The correlation assumes rather natural though not quite trivial forms in two main regions. 
Viz., in the central (moderate scattering angle and energy loss) region, the angular distribution broadens with the increase of the energy loss, 
as if the effective path length of the electron in the target depended on the energy loss (see Sec. \ref{sec:central-region}). 
That should be distinguished from the genuine extension of the particle path length due to the trajectory curvature (``detour''), 
which is negligible at high energy. 
There is also a similarity with the impact-parameter-mediated correlation, with the proviso that the role of the impact parameter is played by the Fourier-reciprocal of the scattering angle on the probability level (see Sec. \ref{sec:transp-eq}).
%One may expect a similar correlation effect to exist also for heavy ionizing particles.

The mentioned similarities pertain to the central domain, 
but the most prominent feature in the correlated distribution function is a spur extending in the large deflection angle and energy loss region. 
It corresponds to a quasifree $ee$ scattering. 
In contrast to the pure kinematical delta function (\ref{dsigmainel-chic2}), 
however, it is significantly smeared by multiple scattering effects (see Sec. \ref{sec:asymptotics}).

Experimental verification of the predicted correlation should be feasible with silicon (lowest-$Z$ semiconductor) targets, by observing $\sim Z^{-1}\approx 10\%$ differences between angular distributions measured at different values of the ionization energy loss (see Figs.~\ref{fig:PlotZ14} and \ref{fig:g-1}b). 
Stronger correlation effects ($\sim Z^{-1}\approx 20\%$, see Figs.~\ref{fig:PlotZ6} and \ref{fig:g-1}b) may become measurable in future with the advent of organic semiconductors \cite{organic-semicond}.

%, or in the hard scattering tail, it also survives in the central region. 
%There, the joint distribution may indeed be approximated by a product of Landau and Moli\`{e}re distributions, but in the latter, 

%The only empirical parameters here are $\chi_a$ and $I$. So, it is impossible to extract therefrom $\Delta\epsilon(b)$.

%We did not take into account the density effect.

\section*{Acknowledgements}

This work was supported in part by the National Academy of Sciences of Ukraine
(projects %0118U100327/
0118U006496, 0120U103567 and 0120U103570). 
%and the Ministry of Education and Science of Ukraine (project 0118U002031).

%\newpage

\appendix

\section{Evaluation of integral (\ref{integral-app})}\label{app:sum}

Integral (\ref{integral-app}) can be evaluated with the demanded NLLA accuracy, 
e.g., by decomposing the Bessel function into power series,
$
J_0(b\chi)=\sum_{n=0}^{\infty}(n!)^{-2}\left(-{b^2\chi^2}/{4}\right)^n
$, 
and integrating termwise:
\begin{eqnarray}\label{}
\int_{\chi_1}^{\infty} \frac{d\chi}{\chi^3} \left[1-J_0(b\chi)e^{- s \chi^2}\right]\qquad\qquad\qquad\qquad\qquad\quad\nonumber\\
=\frac1{2\chi_1^2}-\sum_{n=0}^{\infty}\frac{1}{(n!)^2}\left(-\frac{b^2}{4}\right)^n\int_{\chi_1}^{\infty} d\chi \chi^{-3+2n}  e^{- s \chi^2}.\quad
\end{eqnarray}
In the obtained series, terms with $n\geq2$ have integrands nonsingular at $\chi\to0$, 
and allow to substitute there $\chi_1=0$ within the present accuracy. 
Terms with $n=0$ and $n=1$ have singular integrands, but they can be evaluated exactly:
\begin{eqnarray}\label{n0+n1+series}
\int_{\chi_1}^{\infty} \frac{d\chi}{\chi^3} \left[1-J_0(b\chi)e^{- s \chi^2}\right]\qquad\qquad\qquad\qquad\nonumber\\
=\frac1{2\chi_1^2}
-\int_{\chi_1}^{\infty} \frac{d\chi}{\chi^{3}}  e^{- s \chi^2}
+\frac{b^2}{4}\int_{\chi_1}^{\infty} \frac{d\chi}{\chi} e^{- s \chi^2}\,\,\,\,\nonumber\\
-\sum_{n=2}^{\infty}\frac{1}{(n!)^2}\left(-\frac{b^2}{4}\right)^n\int_{0}^{\infty} d\chi \chi^{-3+2n}  e^{- s \chi^2}\nonumber\\
\equiv\frac1{2\chi_1^2}\left(1-e^{- s \chi_1^2}\right)
+\frac12\left( s +\frac{b^2}{4}\right)E_1\left( s \chi_1^2\right)\nonumber\\
-\frac{ s }{2}\sum_{n=2}^{\infty}\frac{1}{(n-1)n n!}\left(-\frac{b^2}{4 s }\right)^{n},
\end{eqnarray}
where $E_1(z)=\int_z^{\infty}\frac{d\xi}{\xi}e^{-\xi}$ is the exponential integral function \cite{Abr-Steg}. 
At small $ s \chi_1^2$, the terms in the first line may be approximated by
\begin{equation}\label{A3}
\frac1{2\chi_1^2}\left(1-e^{- s \chi_1^2}\right)
\underset{ s \chi_1^2\ll1}\simeq \frac{ s }{2},
\end{equation}
\begin{equation}\label{A4}
E_1\left( s \chi_1^2\right)%=\int_{ s \chi_1^2}^{\infty}\frac{dz}{z}e^{-z} 
\underset{ s \chi_1^2\ll1}\simeq \ln  \frac{1}{s\chi_1^2}-\gamma_{\text{E}}.
\end{equation}
The series in the last line of (\ref{n0+n1+series}), after decomposing the entering fraction into simpler ones as
\[
\frac{1}{(n-1)nn!}=\frac{1}{(n-1)(n-1)!}-\frac{1}{nn!}-\frac{1}{n!},
\]
evaluates in a closed form:
\begin{equation}\label{sum=e+Ein}
\sum_{n=2}^{\infty}\frac{(-z)^{n}}{(n-1)nn!}
=1-e^{-z}-2z+\left(1+z\right)\text{Ein}(z)\geq 0,
\end{equation}
with the complementary exponential integral function \cite{Abr-Steg}
\begin{eqnarray}\label{Ein-def}
\text{Ein}(z)=-\sum_{n=1}^{\infty}\frac{(-z)^n}{nn!}=\int_0^z\frac{1-e^{-\xi}}{\xi}d\xi\nonumber\\
=\gamma_{\text{E}}+\ln z+E_1(z).\qquad\qquad\quad
\end{eqnarray}
Inserting (\ref{A3})--(\ref{sum=e+Ein}) in Eq. (\ref{n0+n1+series}), we are led to Eq. (\ref{appendix-finalresult}).

In limit $ s \to 0$, (\ref{appendix-finalresult}) correctly goes over to (\ref{eq15}),
whereas in limit $b\to 0$, to the readily checkable expression
\begin{eqnarray}\label{}
\int_{\chi_1}^{\infty} \frac{d\chi}{\chi^3} \left(1-e^{- s \chi^2}\right)\qquad\qquad\qquad\qquad\qquad\nonumber\\
= s\, \underset{b\to0}\lim
\left[\ln \frac{2}{b\chi_1}
- \frac12E_1\left(\frac{b^2}{4 s }\right)+\frac12-\gamma_{\text{E}}\right]\nonumber\\
=\frac{ s }2\left(\ln \frac{1}{ s \chi_1^2}+1-\gamma_{\text{E}}\right).\qquad\qquad\qquad\quad
\end{eqnarray}

\section{Correspondence of Eq. (\ref{barlambdaTheta-abc}) with \cite{Jakas} }\label{app:Jakas}

In application to angular dependence of the mean energy loss, it is instructive to compare our expressions (\ref{meanetheta-meanuTheta}), (\ref{barlambdaTheta-abc}) for this quantity with representation
\begin{equation}\label{Jakas}
\bar\epsilon(l,\theta)
= \frac{ n_a l}{ f(l,\theta)}\iint d\sigma_{in}(\chi,\Delta\epsilon)\Delta\epsilon f(l,|\bm\theta-\bm\chi|)
\end{equation}
derived in \cite{Jakas}. 
Eq. (\ref{Jakas}) explicitly involves the inelastic single-scattering differential cross section, 
but under the conditions of multiple scattering, 
usually, little depends on its detail. 

To prove that this is the case here, as well, 
we return to the generic solution (\ref{double-Fourier}), (\ref{fblambda=expJ0e}) of the transport equation in terms of single-scattering cross sections, and insert it to the definition (\ref{meanetheta-meanuTheta}) of the mean energy loss at a fixed $\theta$: 
\begin{eqnarray}\label{B1}
\bar\epsilon(l,\theta)&=&\frac{\int_0^{\infty}d\epsilon\epsilon f(l,\theta,\epsilon)}{f(l,\theta)}\nonumber\\
&=&\frac{p^2}{2m}\frac{l}{2\pi f(l,\theta)} \int_0^{\infty} dbb J_0(b\theta )e^{-l\kappa(b,0)}\frac{\partial\kappa}{\partial s }\bigg|_{ s =0}\nonumber\\
&=&\frac{ n_a l}{2\pi f(l,\theta)}\int_0^{\infty} dbb J_0(b\theta )e^{-l\kappa(b,0)} \nonumber\\
&\,&\qquad\qquad\times\iint d\sigma_{in}(\chi,\Delta\epsilon)\Delta\epsilon J_0(b\chi).
\end{eqnarray}
In the second equality we used identity $\epsilon e^{2mp^{-2}s\epsilon}=\frac{p^2}{2m}\frac{\partial}{\partial s }e^{2mp^{-2}s\epsilon}$, integrated over $s$ by parts, and then performed $\epsilon$ and $s$ integrations. 

If we interchange in (\ref{B1}) the order of integrations,
\begin{eqnarray}\label{}
\bar\epsilon(l,\theta)
=\frac{ n_a l}{  f(l,\theta)}\iint d\sigma_{in}(\chi,\Delta\epsilon)\Delta\epsilon \qquad\qquad \nonumber\\
 \times\int  \frac{d^2b}{(2\pi)^2} e^{i\bm b\cdot(\bm\theta-\bm\chi)-l\kappa(b,0)},
\end{eqnarray}
it is brought to form (\ref{Jakas}).

If instead we adopt the multiple scattering approximation, similar to that of Sec. \ref{sec:transp-eq}, integral (\ref{B1}) breaks in two:
\begin{eqnarray}\label{}
n_a l\iint d\sigma_{in}(\chi,\Delta\epsilon)\Delta\epsilon J_0(b\chi)\qquad\qquad\qquad\qquad\qquad\quad \nonumber\\
=n_a l\iint_{\chi<\chi_1} \!\!\! d\sigma_{in}(\chi,\Delta\epsilon)\Delta\epsilon %\nonumber\\
+\frac{p^2}{2m}2Z\bar\chi_c^2 \int_{\chi_1}^{\infty} \frac{d\chi}{\chi} J_0(b\chi),\nonumber\\
\end{eqnarray}
where in the first term in the right-hand side we have let $J_0(b\chi)\to J_0(0)=1$, while in the second one, employed Rutherford asymptotics (\ref{dsigmainel-chic2}). Utilizing now Eq. (\ref{ioniz-potential-def}) along with identity
\[
\int_{\chi_1}^{\infty} \frac{d\chi}{\chi} J_0(b\chi)\underset{b\chi_1\to\infty}\simeq\ln\frac{2}{b\chi_1}-\gamma_{\text{E}},
\]
the delimiting parameter $\chi_1$ cancels out, and we are left with
\begin{eqnarray}\label{}
n_a l\iint d\sigma_{in}(\chi,\Delta\epsilon)\Delta\epsilon J_0(b\chi)\qquad\qquad\qquad\nonumber\\
= \frac{Z}{m}p^2\bar\chi_c^2 \left(\ln\frac{2p\gamma}{b I_\delta}-\frac{v^2}{2c^2}-\gamma_{\text{E}}\right).
\end{eqnarray}
Insertion thereof to Eq. (\ref{B1}), 
\begin{eqnarray}\label{eqB5}
\bar\epsilon(l,\theta)
=\frac{p^2Z\bar\chi_c^2}{2m f(l,\theta)}\frac{1}{2\pi}\int_0^{\infty} dbb J_0(b\theta )e^{-l\kappa(b,0)}\qquad\quad\nonumber\\
\times\left(2\ln\frac{2p\gamma}{b I_\delta}-\frac{v^2}{c^2}-2\gamma_{\text{E}}\right),\quad
\end{eqnarray}
leads back to Eqs. (\ref{meanetheta-meanuTheta}), (\ref{mean-u(Theta)-def}). 
According to (\ref{intvarphiM/varphiM}), this is expressible completely through $\varphi_M$, without the need to know the differential inelastic cross section $d\sigma_{in}$.

Therefore, in the multiple scattering regime the only parameter of (\ref{eqB5}) sensitive to the inelastic single-scattering cross section is $I_{\delta}$. 
Furthermore, in the ultrarelativistic limit, due to the density effect, 
it expresses through the mean electron density, which is determined by $Z$. 
Correlation effects for $\bar\epsilon(l,\theta)$ discussed herein do not involve $I_{\delta}$ at all, 
and thus are insensitive to detail of $d\sigma_{in}$, except $Z$.

\end{document}